\documentclass[sigconf,noacm]{acmart}
\renewcommand\footnotetextcopyrightpermission[1]{} 
\setcopyright{none}
\copyrightyear{2018}
\acmYear{2018}
\acmDOI{XXXXXXX.XXXXXXX}

\acmConference[Conference acronym 'XX]{Make sure to enter the correct
  conference title from your rights confirmation emai}{June 03--05,
  2018}{Woodstock, NY}
\acmPrice{15.00}
\acmISBN{978-1-4503-XXXX-X/18/06}

\newif\ifconf
\conftrue

\newif\ifEXT
\EXTfalse

\usepackage[utf8]{inputenc} 
\usepackage[T1]{fontenc}    
\usepackage{url}            
\usepackage{booktabs}       
\usepackage{amsfonts}       
\usepackage{nicefrac}       
\usepackage{microtype}      
\usepackage{xcolor}         

\usepackage{wrapfig}

\newif\iftr     
\newif\ifall    
\newif\ifconf   
\newif\ifsq     
\newif\ifnonb   
\newif\iftodos  

\newif\ifsqCAP
\newif\ifsqVS
\newif\ifsqVST
\newif\ifsqEN
\newif\ifsqTIT

\newcommand{\ignore}[1]{}

\sqtrue
\sqCAPtrue
\sqENtrue
\sqVStrue
\sqVSTfalse
\sqTITtrue

\sqfalse
\sqCAPfalse
\sqENfalse
\sqVSfalse
\sqTITfalse

\usepackage{etex}
\usepackage{balance}
\usepackage{epstopdf}
\usepackage{placeins}

\usepackage{graphicx}
\usepackage{float}
\usepackage{dblfloatfix}
\usepackage{multirow}
\usepackage{rotating}
\usepackage{makecell}
\usepackage{tabulary}
\usepackage{parcolumns}
\usepackage{tikz}
\usetikzlibrary{tikzmark}

\usepackage{xpatch}
\expandafter\xpatchcmd
\csname pgfk@/tikz/every picture/.@cmd\endcsname
{\thepage}{\arabic{page}}{}{}

\tikzstyle{comment} = [draw, fill=blue!70, text=white, text width=3cm, minimum height=1cm, rounded corners, align=left, font=\scriptsize]
\tikzstyle{background_alg} = [draw, fill=blue!20, opacity=0.4, inner sep=4pt, rounded corners=2pt]

\usetikzlibrary{shapes}
\usetikzlibrary{plotmarks}
\usetikzlibrary{calc, fit}

\usepackage{enumitem}

\usepackage{amsthm}
\usepackage{amsmath,amsfonts}
\usepackage{mathtools,mathrsfs}

\newtheorem{theorem}{Theorem}[section]

\newtheorem{proposition}[theorem]{Proposition}

\newtheorem{obs}{Observation}

\DeclarePairedDelimiter{\ceil}{\lceil}{\rceil}

\usepackage{soul}
\usepackage{fontawesome}
\usepackage{pifont}
\usepackage{textcomp}
\usepackage{booktabs}
\usepackage{url}
\usepackage{pbox}
\usepackage[normalem]{ulem}
\usepackage[10pt]{moresize}

\usepackage{cleveref}
\usepackage[utf8]{inputenc}
\crefname{section}{§}{§§}
\Crefname{section}{§}{§§}

\ifsqCAP
\usepackage[font={normalfont, scriptsize}]{caption}
\usepackage[font={normalfont, scriptsize}]{subcaption}
\else
\usepackage[font={normalfont}]{caption}
\usepackage[font={normalfont}]{subcaption}
\fi

\newcommand{\vspaceSQ}[1]{\ifsqVS\vspace{#1}\fi}
\newcommand{\vspaceSQT}[1]{\ifsqVST\vspace{#1}\fi}
\newcommand{\enlargeSQ}[1]{\ifsqEN\enlargethispage{\baselineskip}\fi}

\ifsqTIT
\usepackage[compact]{titlesec}
\titlespacing*{\section}{0pt}{3pt}{-1pt}
\titlespacing*{\subsection}{0pt}{0pt}{-3pt}
\titlespacing*{\subsubsection}{0pt}{2pt}{1pt}
\fi

\usepackage{xcolor}
\definecolor{darkgrey}{RGB}{70,70,70}
\definecolor{lightgrey}{RGB}{200,200,200}
\definecolor{lyellow}{RGB}{255,255,100}
\definecolor{llyellow}{RGB}{250,250,180}
\definecolor{lgreen}{RGB}{144,238,144}
\definecolor{raphael_comments}{RGB}{13, 145, 24}

\usepackage[customcolors]{hf-tikz}
\hfsetbordercolor{white}
\hfsetfillcolor{vlgray}

\definecolor{vlgray}{rgb}{0.77 0.77 0.77}
\definecolor{ablack}{rgb}{0.2 0.2 0.2}
\definecolor{vllgray}{rgb}{0.9 0.9 0.9}
\definecolor{bblue}{rgb}{0.7 0.7 0.99}

\usepackage{colortbl}

\usepackage{listings}

\ifsq
\else
\fi

\newcommand{\maciej}[1]{\textcolor{blue}{[Maciej: #1]}}

\newcommand{\cesar}[1]{\textcolor{teal}{[Cesare: #1]}}

\newcommand{\todo}[1]{\noindent\textcolor{red}{[TODO: #1]}}

\definecolor{hlL}{rgb}{0.8 0.8 0.99}

\newcounter{highlight}

\newcounter{hlLR}

\newcounter{hlLIR}

\newcounter{hlLIIR}

\newcounter{Ahighlight}

\usepackage{scalerel,stackengine}
\stackMath
\newcommand\rwh[1]{%
\savestack{\tmpbox}{\stretchto{%
  \scaleto{%
        \scalerel*[\widthof{\ensuremath{#1}}]{\kern-.6pt\bigwedge\kern-.6pt}%
                  {\rule[-\textheight/2]{1ex}{\textheight}}
                              }{\textheight}%
}{0.5ex}}%
\stackon[1pt]{#1}{\tmpbox}%
}

\usepackage[hang,flushmargin]{footmisc}

\DeclarePairedDelimiter\abs{\lvert}{\rvert}

\renewcommand{\epsilon}{\ensuremath\varepsilon}

\renewcommand{\phi}{\ensuremath{\varphi}}

\renewcommand{\arraystretch}{1.0}

\trfalse
\conftrue

\trtrue
\conffalse

\begin{document}

\title{Motif Prediction with Graph Neural Networks}

%


\author{Maciej Besta$^{1\dagger}$, Raphael Grob$^1$, Cesare Miglioli$^2$, 
Nicola Bernold$^1$,\break Grzegorz Kwasniewski$^1$, Gabriel Gjini$^1$,
Raghavendra Kanakagiri$^3$, Saleh Ashkboos$^1$,\break Lukas
Gianinazzi$^1$, Nikoli Dryden$^1$, Torsten Hoefler$^{1\dagger}$}
       \affiliation{\vspace{0.5em}$^1$ETH Zurich\quad\quad
       {$^2$}Research Center for Statistics, University of Geneva\quad\quad
       $^3$UIUC\quad\quad
       $^\dagger$Corresponding authors
}

\pagestyle{plain}

\vspaceSQ{-1em}
\begin{abstract}
\vspaceSQT{-1em}
Link prediction is one of the central problems in graph mining. However, recent
studies highlight the importance of \emph{higher-order network analysis}, where
complex structures called motifs are the first-class citizens.
We first show that existing link prediction schemes fail to effectively predict
motifs. To alleviate this, we establish a general \emph{motif prediction
problem} and we propose several heuristics that assess the chances for a
specified motif to appear. To make the scores realistic, our heuristics
consider -- among others -- \emph{correlations between links}, i.e., the
potential impact of some arriving links on the appearance of other links in a
given motif.  Finally, for highest accuracy, we develop a graph neural network
(GNN) architecture for motif prediction. Our architecture offers vertex
features and sampling schemes that capture the rich structural properties of
motifs.
While our heuristics are fast and do not need any training, GNNs ensure highest
accuracy of predicting motifs, both for dense (e.g., $k$-cliques) and for sparse ones (e.g.,
$k$-stars). We consistently outperform the best available
competitor by more than 10\% on average and up to 32\% in area under the
curve. Importantly, the advantages of our approach over schemes based on
uncorrelated link prediction increase with the increasing motif size
and complexity.
We also successfully apply our architecture for predicting more arbitrary
\emph{clusters} and \emph{communities}, illustrating its potential for graph
mining beyond motif analysis.
\end{abstract}

\maketitle

\section{Introduction and Motivation}
\label{sec:intro}

One of the central problems in graph mining and learning is link
prediction~\cite{lu2011link, al2006link, taskar2004link, al2011survey,
zhang2018link, zhang2020revisiting}, in which one is interested in assessing
the likelihood that a given \emph{pair of vertices} is, or may become, connected.
%
%
%
However, recent works argue the importance of \emph{higher-order graph
organization}~\cite{benson2016higher}, where one focuses on finding and
analyzing small recurring \emph{subgraphs} called \emph{motifs} (sometimes
referred to as \emph{graphlets} or \emph{graph patterns}) instead of individual
links.
Motifs are central to many graph mining problems in computational biology,
chemistry, and a plethora of other fields~\cite{besta2021graphminesuite,
besta2021sisa, cook2006mining, jiang2013survey, horvath2004cyclic,
chakrabarti2006graph, besta2019slim}.
Specifically, motifs are building blocks of different networks, including
transcriptional regulation graphs, social networks, brain graphs,
or air traffic patterns~\cite{benson2016higher}.
There exist many motifs, for example $k$-cliques, $k$-stars, $k$-clique-stars,
$k$-cores, and others~\cite{lee2010survey, jabbour2018pushing, besta2017push}. 
For example, cliques or quasi-cliques are crucial motifs in protein-protein
interaction networks~\cite{bhattacharyya2009mining, li2005interaction}.
A huge number of works are dedicated to motif \emph{counting}, \emph{listing}
(also called \emph{enumeration}), or \emph{checking for the existence} of a
given motif~\cite{besta2021graphminesuite, cook2006mining}.
%
%
However, while a few recent schemes focus on predicting
\emph{triangles}~\cite{benson2018simplicial, nassar2020neighborhood,
nassar2019pairwise}, no works target the problem of \emph{general motif
prediction}, i.e., analyzing whether specified complex structures may
appear in the data.
As with link prediction, it would enable predicting the evolution of data, but
also finding missing structures in the available data. For example, one could
use motif prediction to find probable missing clusters of interactions in
biological (e.g., protein) networks, and use the outcomes to limit the
number of expensive experiments conducted to find missing
connections~\cite{lu2011link, martinez2016survey}.
%

\vspaceSQT{-0.3em}
In this paper, we first (Section~\ref{sec:motif-prediction}) establish and
formally describe a general motif prediction problem, going beyond link
prediction and showing how to predict
higher-order network patterns that will appear in the future (or which
may be missing from the data).
A key challenge is the appropriate \emph{problem formulation}. Similarly to
link prediction, one wants a \emph{score function} that -- for a given vertex
set~$V_M$ -- assesses the chances for a given motif to appear. Still, the
function must consider the combinatorially increased complexity of the problem (compared to
link prediction). 
In general, contrary to a single link, a motif may be formed by an
\emph{arbitrary} set~$V_M$ of vertices, and the number of potential edges
between these vertices can be large, i.e., $O(|V_M|^2)$. 
For example, one may be interested in analyzing whether a \emph{group} of
entities~$V_M$ may become a $k$-clique in the future, or whether a specific
vertex~$v \in V_M$ will become a \emph{hub} of a $k$-star, connecting $v$ to
$k-1$ other selected vertices from~$V_M \setminus \{v\}$. 
This leads to novel issues, not present in link prediction.
For example, what if \emph{some} edges, belonging to the motif being
predicted, already exist? How should they be treated by a score
function? Or, how to enable users to apply their domain knowledge? For example,
when predicting whether the given vertices will form some chemical particle, a
user may know that the presence of some link (e.g., some specific atomic bond)
may increase (or decrease) the chances for forming another bond. Now, how
could this knowledge be provided in the motif score function?
We formally specify these and other aspects of the problem in a general
theoretical framework, and we provide example motif score functions.
We explicitly consider correlations between edges forming a motif, i.e., the
fact that the appearance of some edges may increase or decrease the overall chances of a given motif to appear.

\begin{figure}[h]
\vspaceSQ{-2em}
\centerline{\includegraphics[width=0.45\textwidth]{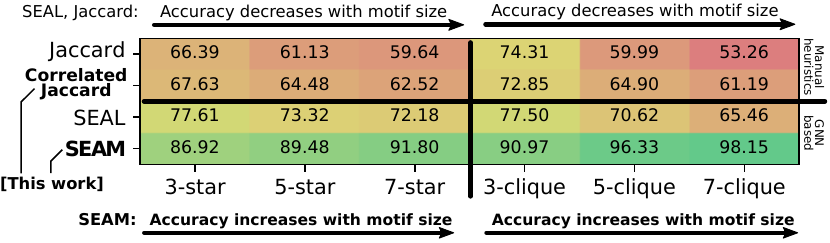}}
\vspaceSQ{-0.5em}
\caption{\textmd{\textbf{Motivating our work (SEAM)}: the accuracy (\%) of predicting
different motifs with SEAM compared to using a state-of-the-art SEAL link
prediction scheme~\cite{zhang2018link, zhang2020revisiting} and a naive one
that does not consider correlations between edges. 
The details of the experimental setup are in Section~\ref{sec:eval} (the dataset is USAir).
\textbf{Importantly: (1)} SEAM outperforms all other methods, \textbf{(2)} the
accuracy of SEAM \emph{increases} with the size ($k$) of each motif, while in
other methods it \emph{decreases}.
}}
\label{fig:intro}
\vspaceSQ{-1em}
\end{figure}

\vspaceSQT{-0.3em}
Then, we develop a learning
architecture based on graph neural networks (GNNs) to further enhance
motif prediction accuracy (Section~\ref{sec:gnns}).
For this, we extend the state-of-the-art SEAL link prediction
framework~\cite{zhang2018link} to support arbitrary motifs. 
For a given motif $M$, we train our architecture on what is the
``right motif surroundings'' (i.e., nearby vertices and edges) that could
result in the appearance of~$M$. Then, for a given set of vertices~$V_M$, the
architecture infers the chances for~$M$ to appear.
The key challenge is to be able to capture the \emph{richness} of different
motifs and their surroundings.
We tackle this with an appropriate selection of \emph{negative}
samples, i.e., subgraphs that resemble the searched motifs but that are not
identical to them.
Moreover, when selecting the size of the ``motif surroundings''
we rely on an assumption also used in link
prediction, which states that only the ``close surroundings'' (i.e., nearby
vertices and edges, 1--2 hops away) of a link to be predicted have a significant
impact on whether or not this link would appear~\cite{zhang2018link,
zhang2020revisiting}. We use this assumption for motifs: as our evaluation
shows, it ensures high accuracy while significantly reducing
runtimes of training and inference (as only a small subgraph is used, instead
of the whole input graph).
We call our GNN architecture \textbf{SEAM}: learning from Subgraphs, Embeddings
and Attributes for \textbf{Motif} prediction\footnote{\scriptsize In analogy to
SEAL~\cite{zhang2018link, zhang2020revisiting}, which stands for ``learning
from Subgraphs, Embeddings, and Attributes for Link prediction''.}.
Our evaluation (Section~\ref{sec:eval}) illustrates the high accuracy of 
SEAM (often more than 90\%), for a variety of graph datasets and motif sizes.

To motivate our work, we now compare SEAM and
a proposed Jaccard-based heuristic that considers link correlations to two
baselines that straightforwardly \emph{use link prediction independently for
each motif link}: a Jaccard-based score and the state-of-the-art SEAL scheme
based on GNNs~\cite{zhang2018link}. We show the results in
Figure~\ref{fig:intro}. The correlated Jaccard outperforms
a simple Jaccard, while the proposed SEAM is better than SEAL.
The benefits
generalize to different graph datasets.
Importantly, we observe that the larger the motif
to predict becomes (larger $k$), \emph{the more advantages our architecture delivers}.
This is because larger motifs provide more room for \emph{correlations between
their associated edges}. Straightforward link prediction based schemes do not
consider this effect, while our methods do, which is why we offer more
benefits for more complex motifs. The advantages of SEAM over the correlated
Jaccard show that GNNs more robustly capture correlations and the structural
richness of motifs than simple manual heuristics. 
Simultaneously, heuristics do not need any training.
%
%
Finally, SEAM also successfully predicts more
arbitrary \emph{communities} or \emph{clusters}~\cite{lee2010survey,
gibson2005discovering, besta2021graphminesuite, besta2021sisa}. They differ
from motifs as they do not have a very specific fixed structure
(such as a star) but simply have
the edge density above a certain threshold. 
SEAM's high accuracy in predicting such structures illustrates its potential
for broader graph mining beyond motif analysis.

\emph{Overall, the key contributions of our paper are (1)
identifying and formulating the motif prediction problem and the associated
score functions, (2) showing how to solve this problem with heuristics and
graph neural networks, and (3) illustrating that graph neural networks can
solve this problem more effectively than heuristics.}

\section{Background and Notation} \label{background}

We first describe the necessary background and notation.

\vspaceSQT{-0.3em}
\textbf{Graph Model}
We model an undirected graph $G$ as a tuple $(V,E)$; $V$ and $E \subseteq V
\times V$ are sets of nodes (vertices) and links (edges); $|V|=n$, $|E|=m$. Vertices are
modeled with integers $1, ..., n$; $V = \{1, ..., n\}$.
$N_v$ denotes the neighbors of $v \in V$; $d(v)$ denotes
the degree of $v$.

\vspaceSQT{-0.3em}
\textbf{Link Prediction}
We generalize the well-known link prediction problem.
%
%
%
Consider two unconnected vertices $u$ and $v$. We assign a \emph{similarity
score} $s_{u,v}$ to them. All pairs of vertices that are not edges receive such
a score and are ranked according to it. The higher a similarity score is, the
``more likely'' a given edge is to be missing in the data or to be created in
the future.  We stress that the link prediction scores are usually not based on
any probabilistic notion (in the formal sense) and are only used to make
comparisons between pairs of vertices in the same input graph dataset.

\vspaceSQT{-0.3em}
There are numerous known similarity scores. First, a large number of scores are
called \emph{first order} because they only consider the neighbors of $u$ and
$v$ when computing $s_{u,v}$. Examples are the \textbf{Common Neighbors} scheme
$s_{u,v}^{CN} = \abs{N_u \cap N_v}$ or the \textbf{Jaccard} scheme
$s_{u,v}^{J} = \frac{\abs{N_u \cap N_v}}{\abs{N_u \cup
N_v}}$~\cite{besta2020communication}.
These schemes assume that two vertices are more likely to be linked if they
have many common neighbors. 
There also exist similarity schemes that
consider vertices not directly attached to $u$ and $v$.
All these schemes can be described using the same formalism of the
\emph{$\gamma$-decaying heuristic} proposed by~\cite{zhang2018link}.
Intuitively, for a given pair of vertices~$(u,v)$, the $\gamma$-decaying
heuristic for $(u,v)$ provides a sum of contributions into the link prediction
score for $(u,v)$ from all other vertices, weighted in such a way that nearby
vertices have more impact on the score. 
%

\vspaceSQT{-0.3em}
\textbf{Graph Neural Networks}
Graph neural networks (GNNs) are a recent class of neural networks for learning
over irregular data such as graphs~\cite{scarselli2008graph,
zhang2019heterogeneous, zhou2020graph, thekumparampil2018attention,
wu2020comprehensive, sato2020survey, wu2020graph, zhang2020deep,
chen2020bridging, cao2020comprehensive}.  There exists a plethora of models and
methods for GNNs; most of them consist of two fundamental parts: (1) an
aggregation layer that combines the features of the neighbors of each node, for
all the nodes in the input graph, and (2) combining the scores into a new
score. 
The input to a GNN is a tuple $G = (A,X)$. The input graph $G$ having $n$
vertices is modeled with an adjacency matrix $A \in \mathbb{R}^{n \times n}$.
The features of vertices (with dimension~$d$) are modeled with a matrix
$X\in \mathbb{R}^{n\times d}$.
%
%

\iftr
\begin{figure*}[b]
\else
\begin{figure}[b]
\fi
\vspaceSQ{-1.5em}
\iftr
  \includegraphics[width=0.75\textwidth]{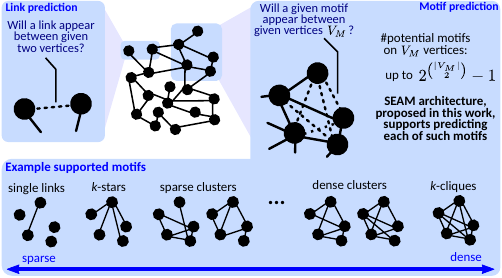}
\fi
\vspaceSQ{-2.5em}
  \caption{\textmd{Illustration of the motif prediction problem and example supported motifs. We provide support for predicting arbitrary motifs.}}
  \vspaceSQ{-1.0em}
  \label{fig:motif-problem}
\iftr
\end{figure*}
\else
\end{figure}
\fi

\iftr
\begin{table*}[t]
\else
\begin{table}[t]
\footnotesize
\fi
\setlength{\tabcolsep}{1.5pt}
\renewcommand{\arraystretch}{0.8}
\vspaceSQ{-1.5em}
\begin{tabular}{ll}\\\toprule  
\textbf{Symbol} & \textbf{Description} \\\midrule
$E_M$ & All edges forming a motif in question; $E_M = E_{M,\mathcal{N}} \cup E_{M,\mathcal{E}}$ \\  
$E_{M,\mathcal{N}}$ & Motif edges that do not yet exist\\
$E_{M,\mathcal{E}}$ & Motif edges that already exist in the data\\
$\overline{E}_{M}$ & Edges not in $E_M$, defined over vertex pairs in $V_M$; $\overline{E}_{M} = \overline{E}_{M,\mathcal{D}} \cup \overline{E}_{M,\mathcal{I}}$ \\
$E_{V_{M}}$ & All possible edges between motif vertices; $E_{V_{M}} = \overline{E}_{M} \cup E_M$ \\
$\overline{E}_{M,\mathcal{D}}$ & Deal-breaker edges; $\overline{E}_{M,\mathcal{D}} = \overline{E}_{M,\mathcal{D},\mathcal{N}} \cup \overline{E}_{M,\mathcal{D},\mathcal{E}}$\\
$\overline{E}_{M,\mathcal{D},\mathcal{N}}$ & Deal-breaker edges that do not exist yet\\
$\overline{E}_{M,\mathcal{D},\mathcal{E}}$ & Deal-breaker edges that already exist\\
$\overline{E}_{M,\mathcal{I}}$ & Non deal-breaker edges in $\overline{E}_{M}$; ``edges that do not matter''\\
$E^*_M$ & ``Edges that matter for the score'': $E^*_M = E_{M} \cup \overline{E}_{M,\mathcal{D}}$\\
$E^{*}_{M,\mathcal{E}}$ & All existing edges ``that matter'': $E^{*}_{M,\mathcal{E}} = E_{M,\mathcal{E}} \cup \overline{E}_{M,\mathcal{D},\mathcal{E}}$\\ 
$E^{*}_{M,\mathcal{N}}$ & All non-existing edges ``that matter'': $E^{*}_{M,\mathcal{N}} = E_{M,\mathcal{N}} \cup \overline{E}_{M,\mathcal{D},\mathcal{N}}$\\ 
\bottomrule
\end{tabular}
\caption{\textmd{Different types of edges used in this work.}}
\label{wrap-tab:1}
\label{tab:edges}
\vspaceSQ{-2em}
\iftr
\end{table*}
\else
\end{table}
\fi
\normalsize

\section{Motif Prediction: Formal Statement and Score Functions}
\label{sec:motif-prediction}

We now formally establish the motif prediction problem.
We define a motif as a pair $M = (V_M, E_M)$. $V_M$ is the set
of \emph{existing} vertices of $G$ that form a given motif ($V_M \subseteq V$). 
$E_M$ is the set of edges of $G$ that form the motif being predicted; some of
these edges may already exist ($E_M \subseteq V_M \times V_M$). 

We make the problem formulation (in \cref{sec:diffs}--\cref{sec:general-f})
\emph{general}: it can be applied to any graph generation process. Using this
formulation, one can then devise specific heuristics that may assume some
details on how the links are created, similarly as is done in link prediction.
Here, we propose example motif prediction heuristics that harness the
Jaccard, Common Neighbors, and Adamic-Adar link scores.
%

We illustrate motif prediction problem and example supported motifs
in Figure~\ref{fig:motif-problem}.

\subsection{Motif Prediction vs.~Link Prediction}
\label{sec:diffs}

We illustrate the motif prediction problem by discussing the differences
between link and motif prediction.
%
%
%
We consider all these differences when proposing specific
schemes for predicting motifs.

\vspaceSQT{-0.3em}
\textbf{(M) There May Be Many Potential New Motifs For a Fixed Vertex Set}
\ 
Link prediction is a ``binary'' problem: for a given pair of unconnected vertices,
there can only be one link appearing. In motif
prediction, the situation is more complex. 
There are many possible motifs to appear between given vertices $v_1, ..., v_k$.
We now state a precise count; the proof is in the appendix.

\vspaceSQT{-0.3em}
\begin{obs} 
Consider vertices $v_1, ..., v_k \in V$. Assuming no edges already connecting
$v_1, ..., v_k$, there are $2^{\binom{k}{2}} - 1$ motifs
(with between 1 and $\binom{k}{2}$ edges) that can appear to connect $v_1, ..., v_k$. 
%

%
\end{obs}

Note that this is the largest possible number, which assumes no previously
existing edges, and permutation dependence, i.e., two motifs that are
isomorphic but have different vertex orderings, are treated as two different
motifs. This enables, for example, the user to be able to distinguish between
two stars rooted at different vertices.  This is useful in, e.g., social
network analysis, when stars rooted at different persons may well have
different meaning.


\vspaceSQT{-0.3em}
\textbf{(E) There May Be Existing Edges}
A link can only appear between \emph{unconnected} vertices. 
Contrarily, a motif can appear and connect
vertices \emph{already} with some edges between them.
%

%

%

\vspaceSQT{-0.3em}
\textbf{(D) There May Be ``Deal-Breaker'' Edges}
There may be some edges, the appearance of which would
make the appearance of a given motif \emph{unlikely} or even \emph{impossible}
(e.g., existing chemical bonds could prevent other bonds).
\iftr
For example, consider a prediction query where one is interested whether a
given vertex set can become connected with a \emph{star} but in such a way that
\emph{none of the non-central vertices are connected to one another}.  Now, if
there is already some edge connecting these non-central vertices, this makes it
impossible a given motif to appear while satisfying the query. 
\fi
We will refer to
such edges as the ``deal-breaker'' edges.

\vspaceSQT{-0.3em}
\textbf{(L) Motif Prediction Query May Depend on Vertex Labeling}
The query can depend on a specific
vertex labeling.  For example, when asking whether a 5-star will connect six
given vertices $v_1, ..., v_6$, one may be interested in
\iftr
\emph{any} 5-star connecting $v_1, ..., v_6$, or 
\fi
a 5-star connecting these vertices in a
\emph{specific way}, e.g., with its center being $v_1$.
%
%
%
%
We enable the user to
specify how edges in~$E_M$
should connect vertices in~$V_M$.

\subsection{Types of Edges in Motifs}

\iftr
\begin{figure*}[t]
\else
\begin{figure}[t]
\fi
  \vspaceSQ{-0.5em}
\iftr
  \includegraphics[width=0.75\textwidth]{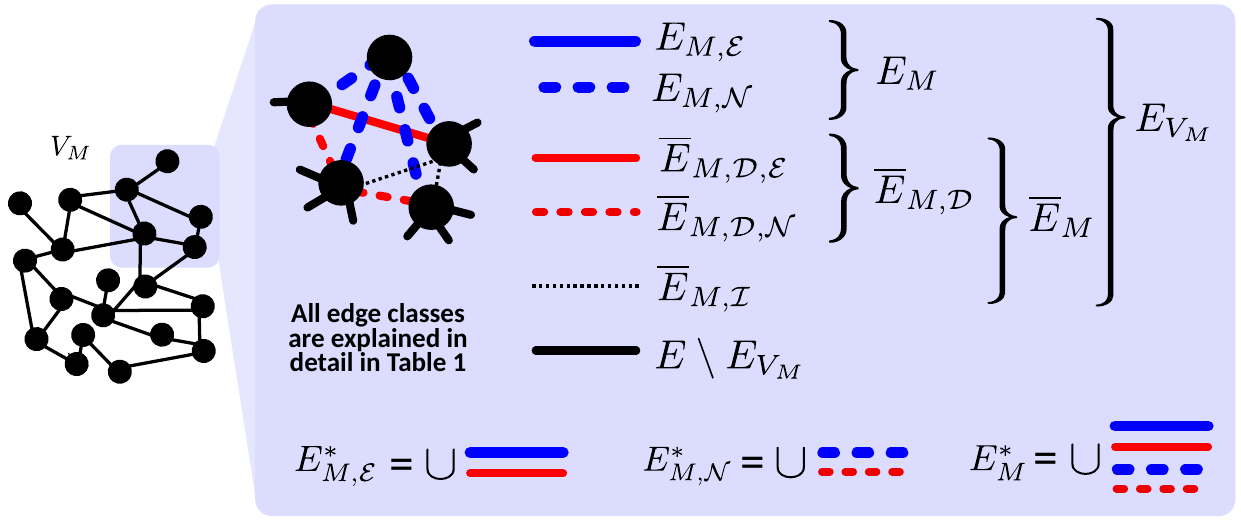}
\fi
\vspaceSQ{-2.5em}
  \caption{\textmd{Illustration of edge types in motif prediction.}}
  \vspaceSQ{-1.0em}
  \label{fig:edges}
\iftr
\end{figure*}
\else
\end{figure}
\fi

We first describe different types of edges related to a motif.
They are listed in Table~\ref{tab:edges} and shown in Figure~\ref{fig:edges}.
First, note that motif edges $E_M$ are a union of two types of motif edges,
i.e., $E_M = E_{M,\mathcal{N}} \cup E_{M,\mathcal{E}}$ where
$E_{M,\mathcal{N}}$ are edges that do not exist in $G$ at the moment of
querying ($\forall_{e \in E_{M,\mathcal{N}}} e \not\in E$; $\mathcal{N}$
indicates ``$\mathcal{N}$on-existing'') and $E_{M,\mathcal{E}}$ are edges that
already exist, cf.~(E) in~\cref{sec:diffs} ($\forall_{e \in E_{M,\mathcal{E}}}
e \in E$; $\mathcal{E}$ indicates ``$\mathcal{E}$xisting'').
Moreover, there may be edges between vertices in $V_M$ which 
do \emph{not} belong to~$M$ (i.e., they belong to $E_{V_{M}} = \{\{i,j\}: i,j \in V_{M} \land i \neq j\}$ but \emph{not} $E_M$). We refer to such edges as
$\overline{E}_{M}$ since $E_{V_{M}} = \overline{E}_{M} \cup E_M$ (i.e., a union
of disjoints sets). Some edges in $\overline{E}_{M}$ may be
deal-breakers (cf.~(D) in~\cref{sec:diffs}), we denote them as
$\overline{E}_{M,\mathcal{D}}$ ($\mathcal{D}$ indicates
``$\mathcal{D}$eal-breaker''). 
Non deal-breakers that are in
$\overline{E}_{M}$ are denoted with $\overline{E}_{M,\mathcal{I}}$
($\mathcal{I}$ indicates ``$\mathcal{I}$nert''). Note that $\overline{E}_{M} =
\overline{E}_{M,\mathcal{D}} \cup \overline{E}_{M,\mathcal{I}}$ and
$\overline{E}_{M} = E_{V_{M}} \setminus E_M$. To conclude, as previously done
for the set $E_M$, we note that $\overline{E}_{M,\mathcal{D}} =
  \overline{E}_{M,\mathcal{D},\mathcal{N}} \cup
  \overline{E}_{M,\mathcal{D},\mathcal{E}}$ where
  $\overline{E}_{M,\mathcal{D},\mathcal{N}}$ are deal-breaker edges that do not
  exist in $G$ at the moment of querying ($\forall_{e \in
  \overline{E}_{M,\mathcal{D},\mathcal{N}}} e \not\in E$; $\mathcal{N}$
  indicates ``$\mathcal{N}$on-existing'') and
  $\overline{E}_{M,\mathcal{D},\mathcal{E}}$ are deal-breaker edges that
  already exist, cf.~(E) in~\cref{sec:diffs} ($\forall_{e \in
  \overline{E}_{M,\mathcal{D},\mathcal{E}}} e \in E$; $\mathcal{E}$ indicates
  ``$\mathcal{E}$xisting''). 
We explicitly consider
$\overline{E}_{M,\mathcal{D},\mathcal{N}}$ because -- even if a given deal-breaker edge
does not exist, but it \emph{does} have a large chance of appearing -- the
motif score should become lower.
%

\subsection{General Problem and Score Formulation}
\label{sec:general-f}

We now formulate a general \emph{motif prediction score}.
Analogously to link prediction, we assign scores to motifs, to be able to
quantitatively assess which motifs are more likely to occur. 
\iftr
Thus, one obtains a tool for analyzing future (or missing) graph structure, by
being able to quantitatively compare different ways in which vertex sets may
become (or already are) connected.
\fi
%
%
Intuitively, we assume that a motif score should be high if the scores of
participating edges are also high. This suggests one could reuse link
prediction score functions.
Full extensive details of score functions, as well as more examples, are in the appendix.

A specific motif score function $s(M)$ will heavily depend on a targeted
problem. In general, we define $s(M)$ as a function of $V_M$ and $E^*_M$; $s(M)
= s(V_M, E^*_M)$.  
Here, $E^{*}_{M} = E_{M} \cup \overline{E}_{M,\mathcal{D}}$ are all the edges
``that matter'': both edges in a motif ($E_{M}$) and the deal-breaker edges 
($\overline{E}_{M,\mathcal{D}}$).
To obtain the exact form of $s(M)$, we harness existing link prediction scores
for edges from $E_M$, when deriving $s(M)$ (details
  in~\cref{sec:ind}--\cref{sec:cor}). When using first-order link prediction
  methods (e.g., Jaccard), $s(M)$ depends on $V_M$ and potential direct
  neighbors. With higher-order methods (e.g., Katz~\cite{katz1953new} or
  Adamic-Adar~\cite{adamic2003friends}), a larger part of the graph that is
  ``around $V_M$'' is considered for computing $s(M)$.
Here, our evaluation (cf.~Section~\ref{sec:eval}) shows that, similarly
to link prediction~\cite{zhang2018link}, it is enough to consider a
small part of $G$ (1-2 hops away from $V_M$) to achieve high prediction
accuracy for motifs.
%

\vspaceSQT{-0.3em}
Still, simply extending link prediction fails to account
for possible \emph{correlations} between edges forming the motif (i.e., edges
  in~$E_M$). Specifically, the appearance of some edges may impact (positively
  or negatively) the chances of one or more other edges in~$E_M$. We provide
  score functions that consider such correlations in~\cref{sec:cor}. 

\subsection{Heuristics with No Link Correlations}
\label{sec:ind}

There exist many score functions for link prediction~\cite{lu2011link,
al2006link, taskar2004link, al2011survey}. Similarly, one can develop
motif prediction score functions with different applications in
mind. 
%
%
As an example, we discuss score functions for a graph that models a set of people. An
edge between two vertices indicates that two given persons know each other. 
For simplicity, let us first assume that there are no deal-breaker edges, thus
$E^*_M = E_M$.
%
%
%
For a set of people $V_M$, we set the score of a given specific
motif $M = (V_M, E_M)$ to be the product of the scores of the associated edges:
$s_{\perp}(M) = \prod_{e \in E_{M,\mathcal{N}}} s(e)$ where $\perp$ denotes the independent aggregation scheme.
Here, $s(e)$ is any link prediction score which outputs into $[0, 1]$ (e.g.,
Jaccard). Thus, also $s_{\perp}(M) \in [0,1]$ by construction. Moreover,
this score implicitly states that $\forall e \in E_{M,\mathcal{E}}$ we
set $s(e) = 1$. Clearly, this does not impact the motif score $s_{\perp}(M)$ as the
edges are already $\mathcal{E}$xisting. Overall, we assume that a motif is more
likely to appear if the edges that participate in that motif are also more
likely.
Now, when using the \textbf{Jaccard Score} for edges, the motif prediction score becomes
$s_{\perp}(M)^{J} = \prod_{e_{u,v} \in E_{M,\mathcal{N}}} \frac{\abs{N_u \cap N_v}}{\abs{N_u \cup N_v}}$.

\vspaceSQT{-0.3em}
To \textbf{incorporate deal-breaker edges},
we generalize the motif score defined previously as
$s_{\perp}^{*}(M) = \prod_{e \in E_{M}} {s(e)} \cdot \prod_{e \in \overline{E}_{M,\mathcal{D}}} \left( 1 - {s(e)}\right)$,
where the product over $E_{M}$ includes partial scores from the edges that
belong to the motif, while the product over $\overline{E}_{M,\mathcal{D}}$
includes the scores from deal-breaker edges.
Here, the larger the chance for a $e$ to appear, the higher its score~$s(e)$
is. Thus, whenever $e$ is a deal-breaker, using $1-s(e)$ has the desired diminishing
effect on the final motif score $s_{\perp}^{*}(M)$.

\subsection{Heuristics for Link Correlations}
\label{sec:cor}

The main challenge is how to aggregate the link predictions taking into account
the rich structural properties of motifs.  Intuitively, using a plain product
of scores implicitly assumes the independence of participating scores. However,
arriving links may increase the chances of other links' appearance in
non-trivial ways. To capture such \emph{positive correlations}, we propose
heuristics based on the \emph{convex linear combination of link scores}.
To show that such schemes consider correlations, we first
(Proposition~\ref{prop:cor-pos}) prove that the product~$P$ of any numbers in
$[0,1]$ is always bounded by the convex linear combination~$C$ of those numbers
(the proof is in the appendix). Thus, our motif prediction scores based on the
convex linear combination of link scores are always at least as large as the
independent products of link scores (as we normalize them to be in~$[0,1]$,
see~\cref{sec:ms-norm}). The difference $(C - P)$ is due to link correlations.
Details are in~\cref{sec:cor-ms-1}.



\begin{proposition}
\label{prop:cor-pos}
	Let $\{x_{1}, ..., x_{n}\}$ be any finite collection of elements from $U = \{x \in \mathbb{R} : 0 \leq x \leq 1\}$. Then, $\forall n \in \mathbb{N}$ we have $\prod_{i=1}^{n} x_{i} \leq \sum_{i=1}^{n} w_{i}x_{i}$,
	where $w_{i} \geq 0 \; \forall i \in \{1, ..., n\}$ and subject to the constraint
	$\sum_{i=1}^{n} w_{i} = 1$. 
\end{proposition}

\vspaceSQT{-0.3em}
For \emph{negative correlations} caused by deal-breaker edges, i.e.,
correlations that lower the overall chances of some motif to appear, we introduce
appropriately normalized scores with a negative sign in the weighted score sum.
The validity of this approach follows from Proposition~\ref{prop:cor-pos} by
noting that $\prod_{i=1}^{n} x_{i} \geq - \sum_{i=1}^{n} w_{i}x_{i}$ under the
conditions specified in the proposition. 
This means that any combination of such negatives scores is always lower than
the product of scores~$P$; the difference~$|C - P|$ again indicates effects between links
not captured by~$P$.
Details are in~\cref{sec:cor-ms-2}.

\subsubsection{Capturing Positive Correlation}
\label{sec:cor-ms-1}


In order to introduce positive correlation, we set the score of a given specific
motif $M = (V_M, E_M)$ to be the convex linear combination of the vector of
scores of the associated edges:

\vspaceSQ{-1em}
\begin{gather}
	\label{eq:cor-s-simple}
	s(M) = f(\mathbf{s(e)}) = \langle \mathbf{w}, \mathbf{s(e)} \rangle 
\end{gather}
\vspaceSQ{-1em}

Here, $f(\mathbf{s(e)}): [0, 1]^{|E_{M}|} \rightarrow [0, 1]$ with $|E_{M}| =
|E_{V_{M}} \setminus \overline{E}_{M}|$ (i.e., not considering either
$\mathcal{I}$nert or $\mathcal{D}$eal-breaker edges). In the weight vector
$\mathbf{w} \in [0, 1]^{|E_{M}|}$, each component $w_{i}$ is larger than zero,
subject to the constraint
$\sum_{i=1}^{|E_{M}|} w_{i} = 1$. Thus, $s(M)$ is a convex linear combination
of the vector of link prediction scores $\mathbf{s(e)}$. 
Finally, we assign a unit score for each existing edge $e \in
E_{M,\mathcal{E}}$. 
%

\vspaceSQT{-0.3em}
Now, to obtain a \textbf{correlated Jaccard score for motifs}, we set a score
for each $\mathcal{N}$on-existing edge $e_{(u,v)}$ as $\frac{\abs{N_u \cap
  N_v}}{\abs{N_u \cup N_v}}$. $\mathcal{E}$xisting edges each receive scores~1. 
  Finally, we set the weights as $\mathbf{w} = \mathbf{1} \frac{1}{|E_{M}|}$, 
assigning the same importance to each link in 
  the motif $M$. This gives
$s(M)^{J} = \frac{1}{|E_{M}|} \;\; \left(\;\;  \sum_{e_{u,v} \in E_{M,\mathcal{N}}}^{} \frac{\abs{N_u \cap N_v}}{\abs{N_u \cup N_v}} + |E_{M,\mathcal{E}}|\;\; \right)$.
%
%
Any choice of $w_{i} > \frac{1}{|E_{M}|}$ places a larger weight on the
$i$-th edge (and lower for others due to the constraint $\sum_{i=1}^{|E_{M}|}
w_{i} = 1$). In this way we can incorporate domain knowledge for the motif of
interest. For example, in Figure~\ref{fig:main-data}, we set $\mathbf{w} = \mathbf{1} \frac{1}{|E_{M,\mathcal{N}}|}$ because of the relevant presence of $\mathcal{E}$xisting edges (each receiving a null score).

\subsubsection{Capturing Negative Correlation}
\label{sec:cor-ms-2}
 
To capture \emph{negative correlation} potentially coming from deal-breaker
edges, we assign negative signs to the respective link scores.
Let $e \in E^{*}_{M} = E_{M} \cup \overline{E}_{M,\mathcal{D}}$. Then we set
$s^{*}_{i}(e) = - {s_{i}(e)}$ if $e \in
\overline{E}_{M,\mathcal{D},\mathcal{N}}$, $\forall i \in \{1, ...,
|E^{*}_{M}|\}$. Moreover, if there is an edge $e \in
\overline{E}_{M,\mathcal{D},\mathcal{E}}$, we have $\mathbf{s^{*}(e)} =
\mathbf{0}$.
Assigning a negative link prediction score to a \emph{potential}
$\mathcal{D}$eal-breaker edge lowers the score of the motif. Setting
$\mathbf{s^{*}(e)} = \mathbf{0}$ when at least one $\mathcal{D}$eal-breaker edge
exists, allows us to rule out motifs which cannot arise. 
We now state a final motif prediction score: 

\vspaceSQ{-1.5em}
\begin{gather}
	\label{eq:cor-s-trans}
	s^{*}(M) = f(\mathbf{s^{*}(e)}) = \max(0, \langle \mathbf{w}, \mathbf{s^{*}(e)} \rangle)
\end{gather}
\vspaceSQ{-1.5em}

Here $s^{*}(M): [0,1]^{|E^{*}_{M}|} \rightarrow [0, 1]$ with $|E^{*}_{M}| \leq
\binom{|V_{M}|}{2}$.  Furthermore, we apply a
rectifier on the convex linear combination of the transformed scores vector
(i.e., $\langle \mathbf{w}, \mathbf{s^{*}(e)} \rangle$) with the rationale that
any negative motif score implies the same impossibility of the motif to appear.
All other score elements are identical to those in Eq.~(\ref{eq:cor-s-simple}).

\subsection{Normalization of Scores for Meaningful Comparisons and General Applicability}
\label{sec:ms-norm}

The motif scores defined so far consider only link prediction scores $s(e)$
with values in $[0, 1]$. Thus, popular heuristics such as Common Neighbors,
Preferential Attachment, and the Adamic-Adar index do not fit into this
framework. 
For this, we introduce a \emph{normalized} score $\nicefrac{s(e)}{c}$ enforcing
$c \geq \ceil{\lVert\mathbf{s(e)} \rVert_{\infty}}$ since the infinity
norm of the vector of scores is the smallest value that ensures the desired
mapping (the ceil function defines a proper generalization as 
$\ceil{\lVert\mathbf{s(e)} \rVert_{\infty}} = 1$ for, e.g., Jaccard~\cite{besta2020communication}). 
%
%
%
To conclude, normalization also enables \emph{the meaningful comparison of
scores of different motifs which may differ in size or in their edge sets
$E_M$}.

%

\begin{figure*}[t]
\centering
\vspaceSQ{-1em}
\includegraphics[width=1.0\textwidth]{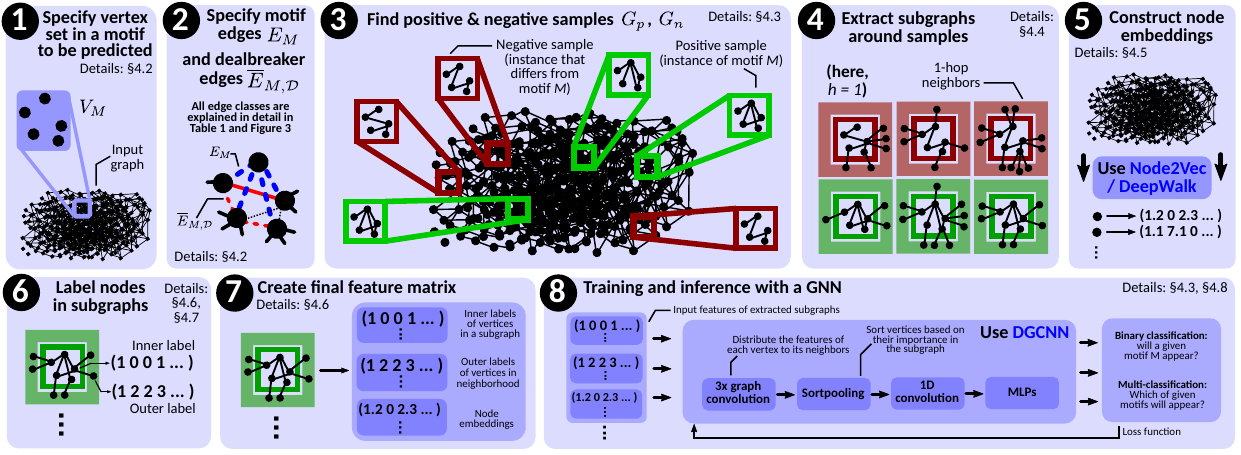}
\vspace{-2em}
\caption{\textmd{High-level overview of SEAM.}}
\label{fig:main-seam}
\vspaceSQ{-1em}
\end{figure*}

\section{SEAM GNN ARCHITECTURE}
\label{sec:gnns}

We argue that one could \emph{use neural networks to learn a heuristic for
motif prediction}. Following recent work on link
prediction~\cite{zhang2018link, zhang2020revisiting}, we use a GNN for this; a
GNN may be able to learn link correlations better than a simple hand-designed
heuristic. Simultaneously, heuristics are still important as they do not
require expensive training.
We now describe a GNN architecture called SEAM (learning from Subgraphs,
Embeddings and Attributes for Motif prediction).
A high-level overview is in~\cref{sec:overview} and in Figure~\ref{fig:main-seam}.

\subsection{Overview}
\label{sec:overview}

Let $M = (V_M, E_M)$ be a motif to be predicted in~$G$. First, we \textbf{extract the
already existing instances of $M$ in $G$}, denoted as $G_p = (V_p, E_p)$;
$V_p \subseteq V$ and $E_p \subseteq E$. 
%
%
We use these instances $G_p$ to \textbf{generate positive samples} for training
and validation. 
%
%
To \textbf{generate negative samples} (details
in~\cref{sec:neg-sampling}), we find subgraphs $G_n = (V_n, E_n)$ that do
\emph{not} form a motif~$M$ (i.e., $V_n = V_M$ and $E_M \nsubseteq E_n$ or
$\overline{E}_{M,\mathcal{D}} \cap E_n \neq \emptyset$).
Then, for each positive and negative sample, consisting of sets of vertices
$V_p$ and $V_n$, we \textbf{extract a ``subgraph around this sample''}, $G_s =
(V_s, E_s)$, with $V_p \subseteq V_s\subseteq V$ and $E_p \subseteq
E_s\subseteq E$, or $V_n \subseteq V_s\subseteq V$ and $E_n \subseteq
E_s\subseteq E$ (details in~\cref{sec:subgraph-e}).
Here, we rely on the insights from SEAL~\cite{zhang2018link} on their $\gamma$-decaying heuristic, i.e.,
it is $G_s$, the
``surroundings'' of a given sample (be it positive or negative), that are important
in determining whether $M$
appears or not.
The nodes of these subgraphs are then \textbf{appropriately labeled} to encode
the structural information (details in~\cref{sec:labeling}). With these labeled
subgraphs, we \textbf{train our GNN}, which classifies each subgraph depending
on whether or not vertices $V_p$ or $V_n$ form the motif~$M$. After
training, we \textbf{evaluate the real world accuracy of our GNN} by
using the validation dataset. 

\subsection{Specifying Motifs of Interest}

The user specifies the motif to be predicted. SEAM provides an interface for
selecting (1) vertices~$V_M$ of interest, (2) motif edges~$E_M$, and (3)
potential deal-breaker edges~$\overline{E}_{M, \mathcal{D}}$.  The user first
picks $V_M$ and then they can specify \emph{any} of up to $2^{\binom{|V_M|}{2}}
- 1$ potential motifs as a target of the prediction.
The interface also enables specifying the vertex ordering, or motif's
permutation invariance.

\subsection{Positive and Negative Sampling} 
\label{sec:neg-sampling}

We need to provide a diverse set of samples to ensure that SEAM works
reliably on a wide range of real data. For the positive samples,
this is simple because the motif to be predicted ($M$) is specified. 
Negative samples are more challenging, because -- for a given motif
-- there are many potential ``false'' motifs.
In general, for each motif $M$, we generate negative samples using
three strategies. (1) We first select positive samples and then remove a few
vertices, replacing them with other nearby vertices (i.e., only a small number
of motif edges are missing or only a small number of deal-breaker edges are
added). Such negative samples closely resemble the positive ones. (2) We randomly
sample $V_M$ vertices from the graph; such negative samples are usually
sparsely connected and do not resemble the positive ones. (3) We select a
random vertex~$r$ into an empty set, and then we keep adding randomly selected
vertices from the union over the neighborhoods of vertices already in the set,
growing a subgraph until reaching the size of~$V_M$; such negative samples may
resemble the positive ones to a certain degree.
The final set of negative samples usually contains about 80\% samples generated
by strategy (1) and 10\% each of samples generated by (2) and (3). This distribution
could be adjusted based on domain knowledge of the input graph (we also
experiment with other ratios). Strategies (2) and (3) are primarily used to avoid overfitting of our model.

{As an example, let our motif $M$ be a 3-clique 
($|V_M|=3$ and $|E_M|=3$). Consider a simple approach of generating negative
samples, in which one randomly samples 3 vertex indices and verifies if there
is a closed 3-clique between them. If we use these samples, in our evaluation
for considered real world graphs, this leads to a distribution of 90\%
  unconnected samples $|E_n|=0$, 9\% samples with $|E_n|=1$ and only about 1\%
  of samples with $|E_n|=2$. Thus, if we train our GNN with this dataset, it
  would hardly learn the difference between open 3-cliques $|E_n|=2$ and closed
  3-cliques $|E_M|=3$. Therefore, we provide our negative samples by ensuring
  that a third of samples are open 3-cliques $|E_n|=2$ and another third of
  samples have one edge $|E_M|=1$. For the remaining third of samples, we use
  the randomly generated vertex indices described above, which are mostly
  unconnected vertices $|E_M|=0$.}

\iftr

For dense subgraphs, the sampling is less straightforward.  Overall, the goal
is to find samples with edge density being either close to, or far away from,
the density threshold of a dense subgraph to be predicted. If the edge density
of the sampled subgraph is lower than the density threshold it becomes a
negative sample and vice versa. The samples chosen further away from the
density threshold are used to prevent overfitting similar to strategies (2) and
(3) from above. For this, we grow a set of vertices~$R$ (starting with a single
random vertex), by iteratively adding selected neighbors of vertices in~$R$
such that we approach the desired density. 

\fi

Overall, we choose equally many positive and negative samples to ensure a
balanced dataset. Furthermore, we limit the number of samples if there are too
many, by taking a subset of the samples (selected uniformly at random). 
The positive and negative samples are split into a training dataset and a
validation dataset. This split is typically done in a $9/1$ ratio. To ensure
an even distribution of all types of samples in these two datasets, we randomly
permute the samples before splitting them.

\subsection{Extracting Subgraphs Containing Samples} 
\label{sec:subgraph-e}

To reduce the computational costs of our GNN, we do not use the entire
graph $G$ as input in training or validation. Instead, we rely on recent
insights on link prediction with GNNs~\cite{zhang2018link,
zhang2020revisiting}, which illustrate that it suffices to provide a subgraph
capturing the ``close surroundings'' (i.e., 1--2 hops away) of the vertices we
want to predict a link between, cf.~Section~\ref{background}.
We take an analogous assumption for motifs (our evaluation confirms the
validity of the assumption). 
%
%
For this, we define the ``surroundings'' of a given motif $M = (V_M, E_M)$.
For  $G=(V,E)$ and $V_M\subseteq V$,
the \emph{$h$-hop enclosing subgraph} $G^h_{V_M}$ is given by the set of nodes
$\{ i\in V\mid \exists  x\in V_M:d(i,x)\leq h\}$.
To actually extract the subgraph, we simply traverse~$G$ 
starting from vertices in~$V_M$, for $h$ hops.
\iftr
%
%
\fi

\subsection{Node Embeddings for More Accuracy}
\label{sec:embedding}

In certain cases, the $h$-hop enclosing subgraph might miss some information
about the motif in question (the details of is missed depend on a specific input graph and
selected motif). To alleviate this, while simultaneously avoiding sampling a
subgraph with large~$h$, we also generate a node embedding $X_E \in
\mathbb{R}^{n \times f}$ which encodes the information about more distant graph
regions using random walks. For this, we employ the established
node2vec~\cite{grover2016node2vec} with the parameters from
DeepWalk~\cite{perozzi2014deepwalk}.
$f$ is the dimension of the low-dimensional vector representation of a node. We
generate such a node embedding once and then only append the embedding vectors
(corresponding to the nodes in the extracted subgraph) to the feature matrix of
each extracted subgraph.
\iftr
We obtain (cf.~\cref{sec:labeling}) 
$X_s= \begin{pmatrix} X_{s_i} & X_{s_E} & X_H & X_L & X_E \end{pmatrix} \in
\mathbb{R}^{s \times (d+2f+2k)}.$
\fi

\iftr
Here, we also extend the SEAL approach called \emph{negative injection} for more
effective embeddings~\cite{zhang2018link, zhang2020revisiting}.
The authors of SEAL observe that if embeddings are constructed using the edge set
containing positive training samples, the GNN would focus on fitting this part
of information.
Thus, SEAL generates embedding based on the edge set containing \emph{also} negative training
samples, which ultimately improves accuracy.
In SEAM, we analogously include all \emph{potential} motif and deal-breaker edges $E^*_M$ of all
training samples to the input graph when generating the node embedding.
\fi
%

\subsection{Node Labeling for Structural Features}
\label{sec:labeling}

In order to provide our GNN with as much structural information as possible, we
introduce two node labeling schemes. These schemes serve as structural learning
features, and we use them when constructing feature matrices of the extracted
subgraphs, fed into a GNN.
Let $s$ be the total number of vertices in the extracted subgraph~$G_s$ and $k$
be the number of vertices forming the motif.  We call the vertices in the
respective samples ($V_p$ or $V_n$) the \emph{inner} vertices since they form a
motif sample. The rest of the nodes in the subgraph~$G_s$ are called
\emph{outer} vertices.

\vspaceSQT{-0.3em}
The first label is simply an enumeration of all the inner vertices. We call this
label the \emph{inner label}. It enables ordering each vertex according to its role in
the motif. For example, to predict a $k$-star, we always assign the
inner label~1 to the star central vertex.
This inner node label gets translated into a one-hot matrix
$H \in \mathbb{N}^{k \times k}$; $H_{ij} = 1$ means that the $i$-th vertex in $V_M$ receives
label~$j$. In order to
include $H$ into the feature matrix of the subgraph, we concatenate $H$ with a
zero matrix $0_{(s-k)k} \in \mathbb{N}^{(s-k) \times k}$, obtaining
$X_H = (H \quad 0_{(s-k)k})^T$.

\vspaceSQT{-0.3em}
The second label is called the \emph{outer} label. The label assigns to each
outer vertex its distances to each inner vertex. Thus, each of the $s-k$ outer
vertices get $k$ labels. The first of these $k$ labels describes the distance
to the vertex with inner label~1. All these outer labels form a labeling matrix
$L \in \mathbb{N}^{(s-k) \times k}$, appended with a zero matrix
$0_{kk}$, becoming $X_L = (0_{kk} \quad L )^T \in
\mathbb{N}^{s \times k}$. 
The final feature matrix~$X_s$ of the respective subgraph~$G_s$ consists of
$X_H$, $X_L$, the subgraph node embedding matrix $X_{E}$ and the subgraph input feature matrix $X_{s_i} \in \mathbb{R}^{s
\times d}$; we have $ X_s= \begin{pmatrix} X_{s_i} & X_{E} & X_H & X_L \end{pmatrix}
\in \mathbb{R}^{s \times (d+f+2k)}$; $d$ is the dimension of the input feature
vectors and $f$ is the dimension of the node embedding vectors.

\subsection{Different Orderings of Motif Vertices}

SEAM supports predicting both motifs where vertices have pre-assigned specific
roles, i.e., where vertices are \textbf{permutation dependant}, and motifs with
vertices that are \textbf{permutation invariant}.
The former enables the user to assign vertices meaningful different structural
roles (e.g., star roots).
%
%
The latter enables predicting motifs where the vertex order does not matter.
For example, in a clique, the structural roles of all involved vertices are
equivalent (i.e., these motifs are vertex-transitive).
%
This is achieved by permuting the inner labels according to the applied vertex
permutation.

\subsection{Used Graph Neural Network Model}
\label{sec:main-gnn}

For our GNN model, we use the graph classification neural network DGCNN~\cite{zhang2018end},
used in SEAL~\cite{zhang2018link, zhang2020revisiting}. We now summarize its
architecture. The first stage of this GNN consist of three graph convolution
layers (GConv). Each layer distributes the vertex features of each vertex to
its neighbors. Then, we feed the output of each of these GConv layers
into a layer called $k$-sortpooling where all vertices are sorted based on
their importance in the subgraph. After that, we apply a standard 1D
convolution layer followed by a dense layer, followed by a softmax layer to get
the prediction probabilities.

\vspaceSQT{-0.3em}
The input for our GNN model is the adjacency matrix of the selected $h$-hop
enclosing subgraph $G^h_{V_s}$ together with the feature matrix $X_s$. With
these inputs, we train our GNN model for 100 epochs. After each epoch, 
to validate the accuracy, we simply generate
$G^h_{V_p}$ and $G^h_{V_n}$ as well as their feature matrix $X_p$ and $X_n$
from our samples in the validation dataset. We know for each set of vertices
$V_p$ or $V_n$, if they form the motif $M$. Thus, we can analyse the accuracy
of our model by comparing the predictions with the original information about
the motifs. Ultimately, we expect our model to predict the set of
vertices $V_p$ to form the motif $M$ and the set of vertices $V_n$ not to form
the motif $M$. 
%

\subsection{Computational Complexity of SEAM}
\label{sec:complexities}

We discuss the time complexity of different parts of SEAM, showing that motif
prediction in SEAM is computationally feasible even for large graphs and
motifs.
Assume that $k$, $t$, and $d$ are \#vertices in a motif, the number of
mined given motifs per vertex, and the maximum degree in a graph, respectively.

First, extracting samples depends on a motif of interest. For example,
\textbf{positive sampling} takes $O(n d^k)$ ($k$-cliques), $O(t m)$
($k$-stars), $O(n d^k)$ ($k$-db-stars), and $O(n d k^3)$ (dense clusters).
These complexities assume mining \emph{all} instances of respective motifs;
SEAM further enables fixing the number of samples to find upfront, which
further limits the complexities.
\textbf{Negative sampling} (of a single instance) takes $O(d k)$ ($k$-cliques),
$O(d)$ ($k$-stars), $O(d + k^2)$ ($k$-db-stars), and $O(n d k^3)$ (dense
clusters).  The complexities may be reduced is the user chooses to fix
sampling counts.
%
%
The \textbf{$h$-hop subgraph extraction}, and inner and outer
\textbf{labeling}, take -- respectively -- $O(k d^{2h})$ and $O(k^2 d^h)$ time
per sample.
%
Finally, \textbf{finding node embeddings} (with Node2Vec) and \textbf{training
as well as inference} (with DGCNN) have complexities as described in detail in
the associated papers~\cite{zhang2018end, grover2016node2vec}; they were
illustrated to be feasible even for large datasets.  

\begin{figure*}[t]
\centering
\vspaceSQ{-1em}
\includegraphics[width=1.0\textwidth]{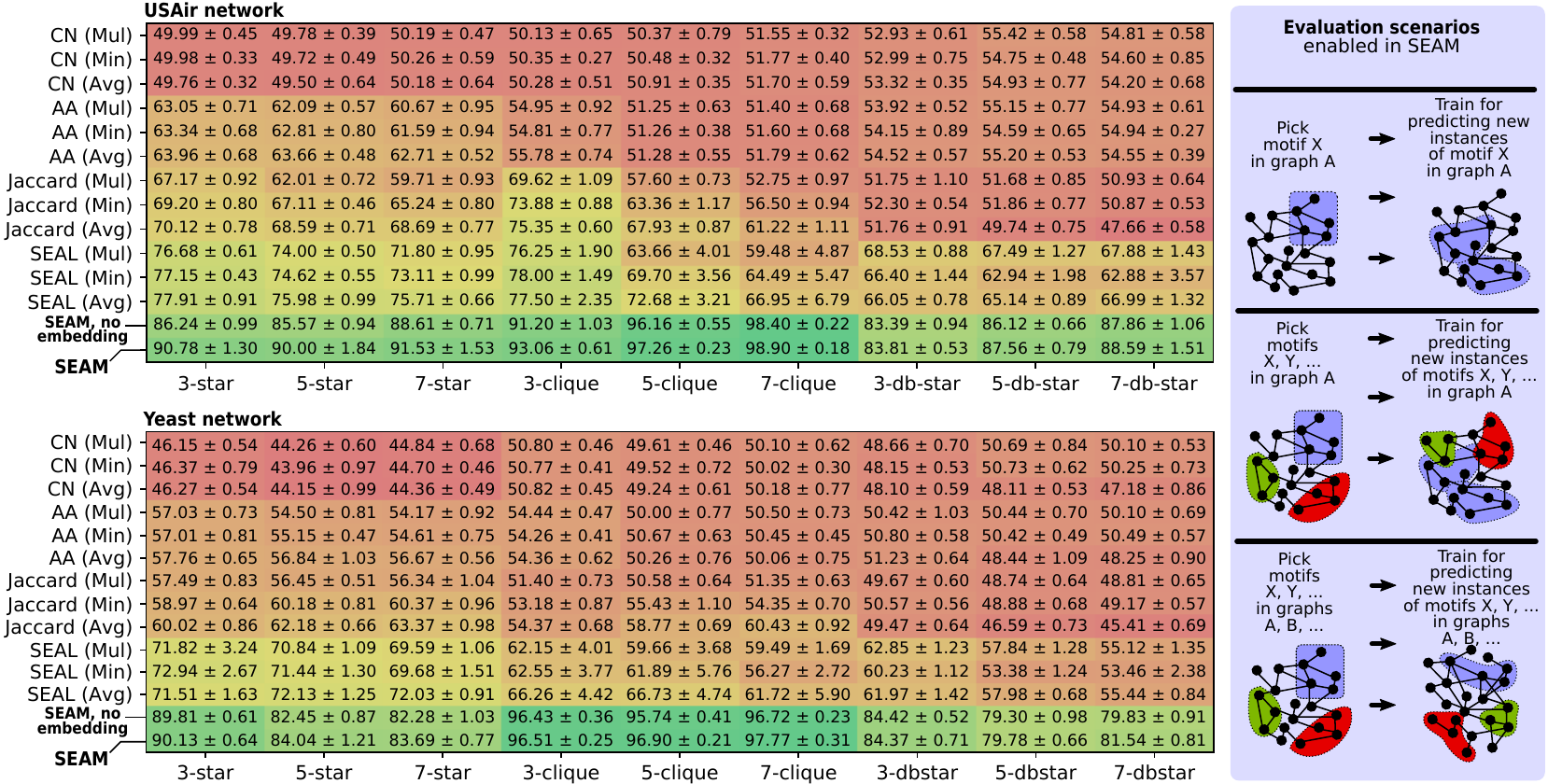}
\vspaceSQ{-2em}
\caption{\textmd{Comparison of different motif prediction schemes; SEAM is the proposed
GNN based architecture. Other baselines use different link prediction schemes
as building blocks; CN stands for Common Neighbors, AA stands for Adamic-Adar.
We use graphs also used by the SEAL link prediction method~\cite{zhang2018link,
zhang2020revisiting}. ``$k$-db-star'' indicate motifs with deal-breaker edges
considered. In the presented data, we predict new instances of a given selected
motif in a given graph dataset.}}
\label{fig:main-data}
\vspaceSQ{-0.5em}
\end{figure*}

\section{Evaluation}
\label{sec:eval}

We now illustrate the advantages of our correlated heuristics and of our
learning architecture SEAM. We feature a representative set of results,
extended results are in the appendix.

\vspaceSQT{-0.3em}
As \textbf{comparison targets}, we use motif prediction based on three link
prediction heuristics (Jaccard, Common Neighbors, Adamic-Adar), and on the GNN
based state-of-the-art SEAL link prediction scheme~\cite{zhang2018link,
zhang2020revisiting}. Here, the motif score is derived using a product of link
scores with no link correlation (``Mul''). We also consider our correlated
heuristics, using  link scores, where each score is assigned
the same importance (``Avg'', $\mathbf{w} = \mathbf{1} \frac{1}{|E_{M,\mathcal{N}}|}$), or
the \emph{smallest} link score is assigned the \emph{highest} importance
(``Min'').  \emph{This gives a total of 12 comparison targets}. We then consider different
variants of SEAM (e.g., with and without embeddings described in~\cref{sec:embedding}). 
More details on the evaluation setting are presented on the right side of Figure~\ref{fig:main-data}.

To assess \textbf{accuracy}, we use AUC (Area Under the Curve), a standard
metric to evaluate the accuracy of any classification model in machine
learning. We also consider a plain fraction of all correct predictions; these
results resemble the AUC ones, see the appendix.

Details of \textbf{parametrization} and \textbf{datasets} are included in the
appendix. In general, we use the same datasets as in the SEAL
paper~\cite{zhang2020revisiting} for consistent comparisons; these are, among
others, Yeast (protein-protein interactions), USAir (airline connections), and
Power (a power grid).
Overall, our current selection of tested motifs covers the whole motif spectrum
in terms of their density: stars (\textbf{very sparse}), communities
(\textbf{moderately sparse} and \textbf{dense}, depending on the threshold),
and cliques (\textbf{very dense}). 

We ensure that the used {graphs match our motivation}, i.e., {they
are either evolving or miss higher order structures} that are then predicted.
For this, we prepare the data so that different edges are removed randomly,
imitating noise.

%

\subsection{SEAM GNN vs.~SEAL GNN vs.~ Heuristics}

\begin{figure}
\vspaceSQ{-0.5em}
\iftr
\centerline{\includegraphics[width=0.8\columnwidth]{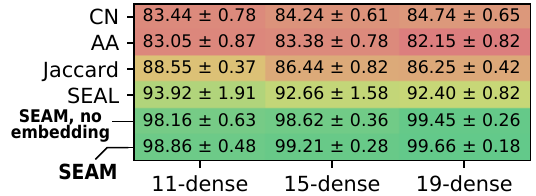}}
\fi
\vspaceSQ{-0.5em}
\caption{\textmd{Comparison of prediction schemes as in Figure~\ref{fig:main-data}
for predicting dense subgraph motif
described in~\cref{sec:neg-sampling}. All link prediction based schemes
use the same motif score. We use the Yeast graph, also used by the SEAL link
prediction method~\cite{zhang2018link,
zhang2020revisiting}.}}
\label{fig:dense-data}
\end{figure}

We compare the accuracy of (1) our heuristics from
Section~\ref{sec:motif-prediction}, (2) a scheme using the SEAL link
prediction, and (3) our proposed SEAM GNN architecture. 
The results for \textbf{$k$-stars}, \textbf{$k$-cliques}, and
\textbf{$k$-db-stars} (for networks USAir and Yeast) are in
Figure~\ref{fig:main-data} while \textbf{clusters} and \textbf{communities} are
analyzed in Figure~\ref{fig:dense-data} (\textbf{``$k$-dense''} indicates a
cluster of $k$ vertices, with at least 90\% of all possible edges present).

\textbf{Behavior and Advantages of SEAM}
\ 
First, in Figure~\ref{fig:main-data}, we observe that the improvement in
accuracy in SEAM almost always scales with the size of the motif. This shows
that SEAM captures correlation between different edges (in larger motifs, there
is more potential correlation between links). Importantly, the advantages and
the capacity of SEAM to capture correlations, also hold in the presence of
\emph{deal-breaker edges} (``$k$-db-star''). Here, we assign links connecting
pairs of star outer vertices as deal-breakers (e.g., 7-db-star is a 7-star with
15 deal-breaker edges connecting its arms with one another).
We observe that the accuracy for $k$-stars with deal-breaker edges is lower
than that for standard \emph{$k$-stars}.
However, SEAM is still the best baseline since it appropriately learns such
edges and their impact on the motif appearance.
The results in Figure~\ref{fig:dense-data} follow similar trends to those in
Figure~\ref{fig:main-data}; SEAM outperforms all other baselines. Its accuracy
also increases with the increasing motif size.
\emph{Overall, SEAM significantly outperforms both SEAL and heuristics in
accuracy, and is the only scheme that captures higher-order characteristics,
i.e., its accuracy increases with the amount of link correlations.}

\textbf{Behavior and Advantages of Heuristics}
\ 
While the core result of our work is the superiority of SEAM in accuracy,
our correlated heuristics (``Avg'', ``Min'') also to a certain degree improve the motif
prediction accuracy over methods that assume link independence (``Mul'').
This behavior holds often in a statistically significant way,
cf.~Jaccard results for 3-cliques, 5-cliques, and 7-cliques. In several cases,
the differences are smaller and fall within the standard deviations of
respective schemes.
Overall, we observe that $AUC_{Mul}< AUC_{Min}<AUC_{Avg}$
(except for $k$-db-stars). This
shows that different aggregation schemes have different capacity in capturing
the rich correlation structure of motifs. In particular, notice that ``Min'' is
by definition (cf.~Proposition~\ref{prop:cor-pos}) a lower bound of the score
$s(M)$ defined in~\cref{sec:cor-ms-1}. This implies that it is the smallest
form of correlation that we can include in our motif score given the convex
linear combination function proposed in~\cref{sec:cor-ms-1}. 

The main advantage of heuristics over SEAM (or SEAL) is that \emph{they do not
require training, and are thus overall faster}.
For example, to predict 100k motif samples, the heuristics take around 2.2
seconds with a standard deviation of 0.05 seconds, while SEAM has a mean
execution time (including training) of 1280 seconds with a standard deviation
of 30 seconds.
Thus, we conclude that heuristics could be preferred over SEAM when training
overheads are deemed too high, and/or when the sizes of considered motifs
(and thus the amount of link correlations) are small.

%

Interestingly, the Common Neighbors heuristic performs poorly. This is due to
the similar neighborhoods of the edges that have to be predicted. The high
similarity of these neighborhoods is caused by our subgraph extraction strategy
discussed in Section~\ref{sec:subgraph-e}, where we select the existing motif
edges of the positive samples in such a way as to mimic the edge structure of
the negative samples. These results show also that different heuristics do not
perform equally with respect to the task of motif prediction and further
studies are needed in this direction.

The accuracy benefits of SEAM over the best competitor
(SEAL using the ``Avg'' way to compose link prediction scores into motif
prediction scores) range from 12\%  to almost 32\%. This difference is even
larger for other methods; it is because there comparison targets cannot
effectively capture link correlations in motifs.  
This result shows that the
edge correlation in motifs is important to accurately predict a motif's
appearance, and that it benefits greatly from being learned by a neural
network.

\subsection{Additional Analyses}

\iftr
The only difference in this dataset is the slight drop in accuracy for bigger
stars and stars with deal-breaker edges. We conjecture this is because (1) this
dataset has many challenging negative samples (\ref{sec:neg-sampling}) for
bigger motifs, and (2) the neighborhoods of negative and positive samples being
almost indistinguishable.
\fi
We also consider a Power graph, see Figure~\ref{fig:full-power}. This graph
dataset is very sparse with a very low average vertex degree of 2.7 (see the
appendix for dataset details). SEAM again offers the best accuracy.

\iftr
\begin{figure*}[t]
\else
\begin{figure}[h]
\fi
\centering
\vspaceSQ{-0.5em}
\iftr
\includegraphics[width=0.75\textwidth]{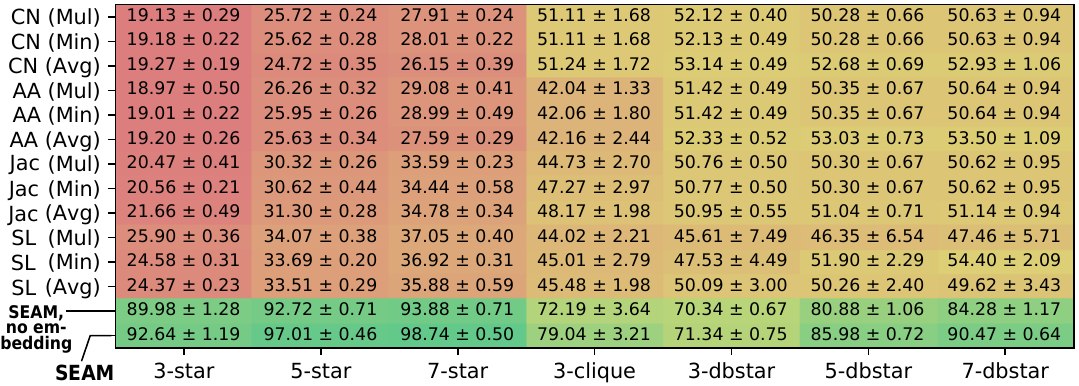}
\fi
\vspaceSQ{-2em}
\caption{\textmd{Comparison of different motif prediction schemes for a very sparse
Power (power grid) graph. CN: Common Neighbors, AA: Adamic-Adar, Jac: Jaccard,
SL: SEAL. ``$k$-db-star'': motifs with deal-breaker edges
considered.}}
\label{fig:full-power}
\vspaceSQ{-0.5em}
\iftr
\end{figure*}
\else
\end{figure}
\fi

\iftr
This result clearly shows very low accuracy of SEAL and other motif scores if
there are just a few vertices in the neighborhood of the motif. The prediction
accuracy for {$k$-stars} with deal-breaker edges is significantly better. This
is caused by the properties of the positive samples discussed in
Section~\ref{sec:neg-sampling}. The prediction task of these positive samples
boils down to predicting one motif edge, which has to be added, and several
deal-breaker edges, that cannot appear. Due to the sparsity of the motif
neighborhood, these deal-breaker edges are often predicted correctly to not
appear, which significantly increases the prediction strength of SEAL and all
the other motif scores.
\fi

We also analyze the impact of additionally using Node2Vec node embeddings (cf.~\cref{sec:embedding}). Interestingly, it consistently
(by 0.2 -- 4\%) improves the accuracy while simultaneously \emph{reducing the
variance} in most cases by around 50\% for {cliques} and dense clusters.

We also consider other aspects, for example, we vary the number of
existing edges, and even eliminate all such edges; the results
follow similar patterns to those observed above.

SEAM's running times heavily depend
on the used model parameters. A full SEAM execution on the Yeast graph dataset
with 40,000 training samples and 100 training epochs does typically take
between 15--75 minutes (depending on the complexity of the motif, with
stars and dense clusters being the fastest and slowest to process,
respectively). The used hardware configuration includes an Intel 6130 @2.10GHz
with 32 cores and an Nvidia V100 GPU; details are in the appendix.

Other analyses are in the appendix, they include varying the used labeling
schemes, training dataset sizes, learning rates, epoch counts, or sizes of
enclosing subgraphs.

\section{Related Work}

Our work is related to various parts of data mining and learning.
\iftr
First, we generalize the established link prediction problem~\cite{lu2011link,
al2006link, taskar2004link, al2011survey, zhang2018link, zhang2020revisiting, cook2006mining, jiang2013survey, horvath2004cyclic,
chakrabarti2006graph}
into arbitrary higher-order structures, considering inter-link correlations,
and providing prediction schemes based on heuristics and GNNs.
Next, many works exist on listing, counting, or finding different patterns
(also referred to as motifs, graphlets, or
subgraphs)~\cite{besta2021graphminesuite, besta2021sisa, chakrabarti2006graph,
washio2003state, lee2010survey, rehman2012graph, gallagher2006matching,
ramraj2015frequent, jiang2013survey, aggarwal2010managing, tang2010graph,
leicht2006vertex, besta2017push, liben2007link, ribeiro2019survey, lu2011link,
al2011survey, bron1973algorithm, cazals2008note, DBLP:conf/isaac/EppsteinLS10,
DBLP:journals/tcs/TomitaTT06, danisch2018listing, jabbour2018pushing}. Our work
enables predicting any of such patterns. Moreover, SEAM can use these schemes as
subroutines when mining for specific samples.
\fi
Third, different works analyze the temporal aspects of
motifs~\cite{kovanen2013temporal, torkamani2017survey}, for example by
analyzing the temporal dynamics of editor
interactions~\cite{jurgens2012temporal}, temporal dynamics of motifs in general
time-dependent networks~\cite{paranjape2017motifs, kovanen2011temporal},
efficient counting of temporal motifs~\cite{liu2019sampling}, predicting
triangles~\cite{benson2018simplicial, nassar2019pairwise}, or using motif
features for more effective link predictions~\cite{abuoda2019link}. However,
none of them considers prediction of general motifs. 
\iftr
Moreover, there exists an extensive body of work on graph processing and
algorithms, both static and dynamic (also called temporal, time-evolving, or
streaming)~\cite{besta2019demystifying, besta2019practice, sakr2020future,
Besta:2015:AIC:2749246.2749263, gianinazzi2018communication, besta2020high,
solomonik2017scaling, besta2020substream, besta2017slimsell, ediger2010massive, ediger2012stinger,
bulucc2011combinatorial, kepner2016mathematical, mccoll2013new, madduri2009faster, kepner2016mathematical}.
\fi
%
%
Still, they do not consider prediction of motifs.

Finally, GNNs have recently become a subject of intense
research~\cite{wu2020comprehensive, zhou2020graph, scarselli2008graph,
zhang2020deep, chami2020machine, hamilton2017representation,
bronstein2017geometric, kipf2016semi, gianinazzi2021learning, bronstein2021geometric,
wu2020comprehensive, zhou2020graph, zhang2020deep, chami2020machine,
hamilton2017representation, bronstein2017geometric, zhang2019graph}. In this
work, we use GNNs for making accurate predictions about motif appearance.
While we pick DGCNN as a specific model to implement SEAM, other GNN models can
also be used; such an analysis is an interesting direction for future work.
\iftr
An interesting venue of future work would be harnessing GNNs for other graph
related tasks, such as compression~\cite{besta2019slim, besta2018survey,
besta2018log}.
\fi
We implement SEAM within the Pytorch Geometric GNN framework. Still,
other GNN frameworks could also be
used~\cite{li2020pytorch, fey2019fast, zhu2019aligraph,
wu2021seastar, hu2020featgraph,  
zhang2020agl}. 
\iftr
An interesting line of work would be to
implement motif prediction using the
serverless paradigm~\cite{jonas2019cloud, mcgrath2017serverless, baldini2017serverless, copik2020sebs},
for example within one of recent dedicated serverless engines~\cite{thorpe2021dorylus}.
\fi

\section{Conclusion \& Discussion}
\label{sec:conclusion}

Higher-order network analysis is an important approach for mining
irregular data. Yet, it lacks methods and tools for predicting the
evolution of the associated datasets.
For this, we establish a problem of predicting general complex graph structures
called motifs, such as cliques or stars. We illustrate its differences to
simple link prediction, and then we propose heuristics for motif prediction
that are invariant to the motif size and capture potential correlations between
links forming a motif.  Our analysis enables incorporating domain
knowledge, and thus -- similarly to link prediction -- it can be a foundation
for developing motif prediction schemes within specific
domains. 

While being fast, heuristics leave some space for improvements in
prediction accuracy. To address this, we develop a graph neural network (GNN)
architecture for predicting motifs. We show that it outperforms the state of
the art by up to 32\% in area under the curve, offering excellent accuracy, which \emph{improves} with the growing
size and complexity of the predicted motif.
We also successfully apply our architecture to predicting more arbitrarily
structured clusters, indicating its broader potential in mining irregular data.

\if 0\maciej{cool ideas but dangerous to include them}
\vspace{-0.3em}
The limitations of our work, pave the way to future challenging contributions.
Indeed we are able to show (see Figure \ref{fig:main-data}) that well-known
heuristics do not perform equally with respect to the task of motif prediction.
However, given the significant computational advantage that they possess, an
overall evaluation of existing heuristics would be of practical interest
especially to define patterns of performance depending possibly on the specific
structure of the motif. In this direction, learning the weights of the general
motif score $s^{*}(M)$ can be a successful strategy to solve the trade-off
between accuracy and computational burden. Moreover, other fruitful insights
can come from extending the Maximum Likelihood Methods (e.g. \cite{lu2011link})
for motif prediction. This because it could provide a model based estimation of
  relevant parameters and, more interestingly, the variability of these
  estimates to potentially define a confidence interval for our motif score
  predictions.
\fi

\iftr

\vspace{1em}
\textbf{Acknowledgements}
We thank Hussein Harake, Colin McMurtrie, Mark Klein, Angelo Mangili, and the
whole CSCS team granting access to the Ault and Daint machines, and for their
excellent technical support. We thank Timo Schneider for immense help with
computing infrastructure at SPCL.
This research received funding from Google European Doctoral Fellowship,
Huawei, and the European Research Council (ERC) under the European Union's Horizon
2020 programme (grant agreement DAPP, No.~678880). 
%

\fi

{
\bibliographystyle{abbrv}
\bibliography{references}
}

\newpage


\appendix


\section*{Appendix A: Proofs}

We recall the statement of \textbf{Observation 1} in Section~\ref{sec:diffs}:
 
\textit{Consider vertices $v_1, ..., v_k \in V$. Assuming no edges already connecting
    $v_1, ..., v_k$, there are $2^{\binom{k}{2}} - 1$ motifs
    (with between 1 and $\binom{k}{2}$ edges) that can appear to connect $v_1, ..., v_k$. 
}

\vspaceSQ{-0.5em}
\begin{proof}
    We denote as $E_{k} = \{\{i,j\}: i,j \in V_{k} \land i \neq j\}$ the edge set of the undirected subgraph
    $(V_{k},E_{k})$ with $V_{k} \subseteq V$. The number of all possible edges between $k$
    vertices is $\vert E_{k} \vert = \binom{k}{2}$. Any subset of
    $E_{k}$, with the exception of the empty set, defines a motif. Thus the set
    of all possible subsets (i.e., the \textit{power set} $\mathcal{P}$) of
    $E_{k}$ is the set of motifs. Then, since $\vert \mathcal{P}(E_{k})\vert =
    2^{\binom{k}{2}}$, we subtract the empty set (which we consider as an
    invalid motif) from the total count to obtain the desired result.
    \end{proof}

\vspaceSQ{-0.5em}
We recall the statement of \textbf{Proposition \ref{prop:cor-pos}} in Section~\ref{sec:cor}:

\textit{Let $\{x_{1}, ..., x_{n}\}$ be any finite collection of elements from $U = \{x \in \mathbb{R} : 0 \leq x \leq 1\}$. Then, $\forall n \in \mathbb{N}$ we have $\prod_{i=1}^{n} x_{i} \leq \sum_{i=1}^{n} w_{i}x_{i}$,
    where $w_{i} \geq 0 \; \forall i \in \{1, ..., n\}$ and subject to the constraint
    $\sum_{i=1}^{n} w_{i} = 1$.}

\vspaceSQ{-0.5em}
\begin{proof}
    We start by noticing that $\prod_{i=1}^{n} x_{i} \leq \min\{x_{1}, ...,
    x_{n}\}$. This is trivial to verify if $\exists \; x_{i} = 0$ for $i \in \{1, ..., n\}$. Otherwise, it can be shown by contradiction: imagine that $\prod_{i=1}^{n} x_{i} > \min\{x_{1}, ...,
    x_{n}\}$. We know that $U$ is closed with respect to the product (i.e., $\prod_{i=1}^{n} x_{i} \in U \; \forall \;n \in \mathbb{N}$). Then, we can divide both sides by $\min\{x_{1}, ...,
    x_{n}\}$, since we ruled out the division by zero, to obtain $\prod_{i=1}^{n-1} x_{i} > 1$. This implies $\prod_{i=1}^{n-1} x_{i} \notin U$, which contradicts that $U$ is closed to the product. For the right side of the original statement, we know by definition that $x_{i} \geq \min\{x_{1}, ...,
    x_{n}\} \; \forall i \in \{1, ..., n\}$. Since $w_{i} \geq 0 $, we can also write that $w_{i} x_{i} \geq w_{i}\min\{x_{1}, ..., x_{n}\} \; \forall i \in \{1, ..., n\}$. Thus,
    since $U$ is an ordered set, we can state that $\sum_{i=1}^{n} w_{i}x_{i} \geq \sum_{i=1}^{n} w_{i} \min\{x_{1}, ..., x_{n}\}$. But then, since $\sum_{i=1}^{n} w_{i} \min\{x_{1}, ..., x_{n}\} =
    \min\{x_{1}, ..., x_{n}\}$, we conclude that $\min\{x_{1}, ..., x_{n}\} \leq
    \sum_{i=1}^{n} w_{i}x_{i}$. This ends the proof thanks to the transitive
    property.
\end{proof}

\vspaceSQ{-0.5em}
We also justify some \textbf{complexity bounds} from~\cref{sec:complexities}.
Mining $k$-stars is independent of $k$. To find a $k$-star at a given node $x$,
one chooses $k-1$ random nodes of $x$. This has a complexity of $O(k+d(x))$. If
$k$ is larger than $d(x)$, there is no star and one can skip the node in
$O(1)$. Otherwise, it takes $O(d(x))$ to extract a star.  Thus, for a given
star, we extract $t$ samples in $O(t d(x))$. Summing over all nodes is hence
$O(t m)$.
Next, the bounds for cliques and clusters are straightforward.
Finally, for the enclosing subgraph extraction and labeling, we do a BFS
starting from each of the $k$ motif vertices that visits all $h$-hop neighbors.
Each node has at most $d$ neighbors, hence there is at most $kd^h$ nodes in the
$h$-hop neighborhood that need to be visited.  But, as BFS is also linear in
the number of edges, the complexity is $O(kd^{2h}$).  The inner labels are a
one-hot-encoding of the motif vertices, which can be produced in $O(k^2 d^h)$
time. The outer labels are the distances to the motif nodes, which can be
computed at the same time as the BFS traversal for the extraction, so it is the
overhead of $O(k^2 d^h)$.

\section*{Appendix B: Details of Datasets}

In this section, we provide additional details on the various datasets that we used.
We selected networks of different origins (biological, engineering, transportation),
with different structural properties (different sparsities and skews in degree distributions).

USAir~\cite{batagelj2006pajek} is a graph with 332 vertices and 2,126 edges representing US cities and the airline connections between them. The vertex degrees range from 1 to 139 with an average degree of 12.8.
%
%
Yeast~\cite{mering2002comparative} is a graph of protein-protein interactions in yeast with 2,375 vertices and 11,693 edges.
The vertex degrees range from 1 to 118 with an average of 9.8.
%
Power~\cite{watts1998collective} is the electrical grid of the Western US with 4,941 vertices and 6,594 edges.
The vertex degrees range from 1 to 19 with an average degree of 2.7.

\section*{Appendix C: SEAM Model Parameters}

\iftr
We now discuss in more detail the selection of the SEAM model parameters.
\fi

\subsection*{Choosing learning rate and number of epochs}

\iftr 
We first describe how we tune the hyperparameters for our motif
prediction framework. 
\fi
To find the optimal learning rate for SEAM we
try different learning rates as shown in Figures~\ref{fig:lr-50},
\ref{fig:lr-100} and~\ref{fig:lr-150}. The associated hyperparameters are
highly dependent on the specific motif to be predicted and on the used dataset.
As an example, we analyze the hyperparameters for {$k$-stars} and
{$k$-cliques} on the USAir graph dataset. The plots show that there is a
sweet spot for the learning rate at 0.001-0.002. Any value below that rate is
too small and our model cannot train its neural network effectively, while for
the values above that, the model is unable to learn the often subtle
differences between hard negative samples and positive samples. The number of
epochs of the learning process can be chosen according to the available
computational resources of the user.

\iftr
\begin{figure*}[t]
\else
\begin{figure}[h]
\fi
\vspaceSQ{-1em}
\iftr
\centerline{\includegraphics[width=0.9\textwidth]{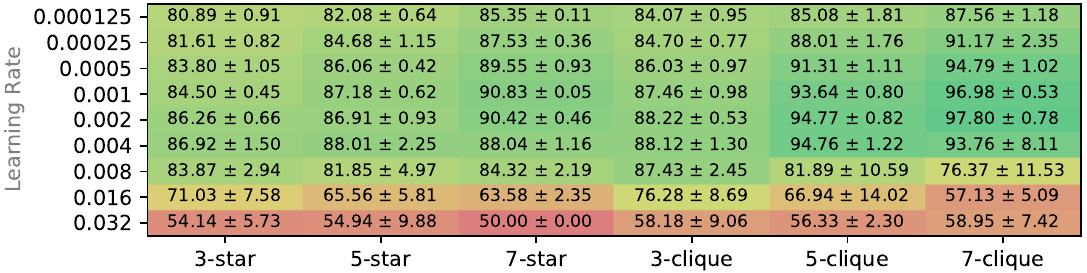}}
\fi
\vspaceSQ{-1em}
\caption{\textmd{AUC-Score comparison for different learning rates on {USAir} graph.
\iftr
SEAM parameters: proposed labels enabled, proposed embedding disabled. 
\fi
Number of epochs = 50, training dataset size = 100,000.}}
\label{fig:lr-50}
\vspaceSQ{-1em}
\iftr
\end{figure*}
\else
\end{figure}
\fi

\iftr
\begin{figure*}[t]
\else
\begin{figure}[h]
\fi
\vspaceSQ{-1em}
\iftr
\centerline{\includegraphics[width=0.9\textwidth]{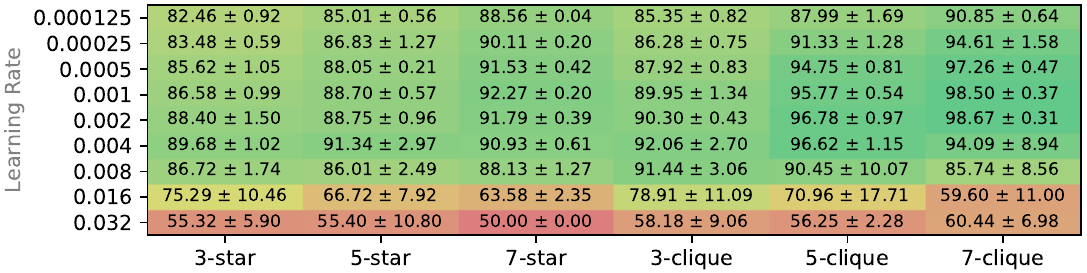}}
\fi
\vspaceSQ{-1em}
\caption{\textmd{AUC-Score comparison for different learning rates on {USAir} graph.
\iftr
SEAM parameters: proposed labels enabled, proposed embedding disabled. 
\fi
Number of epochs = 100, training dataset size = 100,000.}}
\label{fig:lr-100}
\vspaceSQ{-1em}
\iftr
\end{figure*}
\else
\end{figure}
\fi

\iftr
\begin{figure*}[t]
\else
\begin{figure}[h]
\fi
\vspaceSQ{-1em}
\iftr
\centerline{\includegraphics[width=0.9\textwidth]{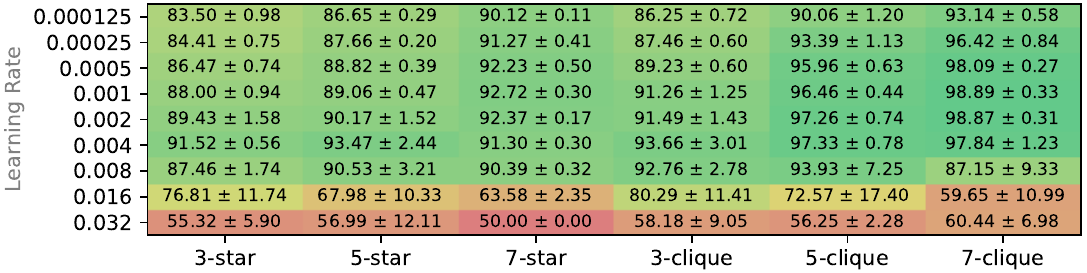}}
\fi
\vspaceSQ{-1em}
\caption{\textmd{AUC-Score comparison for different learning rates on {USAir} graph.
\iftr
SEAM parameters: proposed labels enabled, proposed embedding disabled. 
\fi
Number of epochs = 150, training dataset size = 100,000.}}
\label{fig:lr-150}
\vspaceSQ{-1em}
\iftr
\end{figure*}
\else
\end{figure}
\fi

\begin{figure*}[t]
\vspaceSQ{-1em}
\iftr
\centerline{\includegraphics[width=1.0\textwidth]{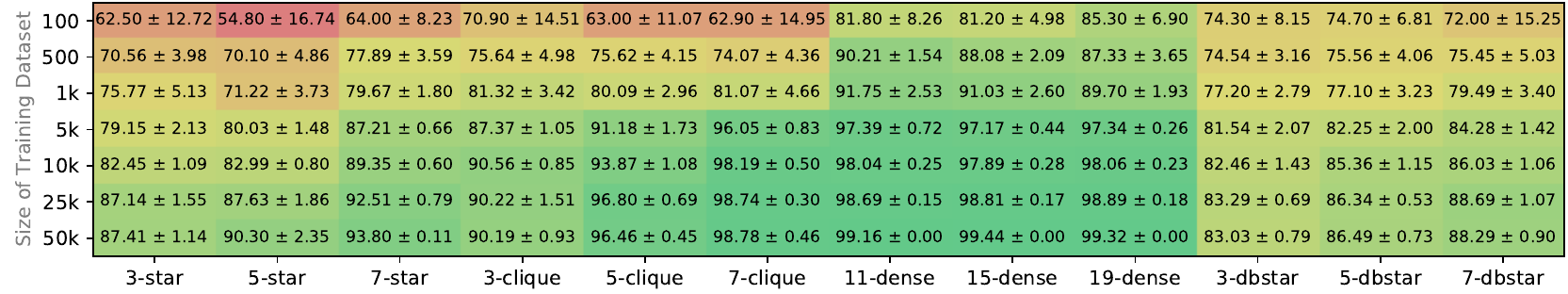}}
\fi
\vspaceSQ{-1em}
\caption{\textmd{AUC-Score comparison for different training dataset sizes on \textbf{USAir} graph.
\iftr
SEAM parameters: proposed labels enabled, proposed embedding enabled.
\fi
Learning rate = 0.002, number of epochs = 100.}}
\label{fig:train-size-usair}
\vspaceSQ{-1em}
\end{figure*}

\subsection*{Analysis of different training dataset sizes}

\iftr
We also analyze the effect of different training dataset sizes on the
prediction strength of SEAM. 
We want to assess the smallest number of samples
that still ensures an effective learning process.
\fi
Figure~\ref{fig:train-size-usair} shows the different accuracy results of SEAM,
for different motifs and training dataset sizes. We observe that the accuracy
  strongly depends on the motif to be predicted. For example, a dense subgraph
  can be predicted with high accuracy with only 100 training samples. On the
  other hand, prediction accuracy of the {5-star} motif improves
  proportionally to the amount of training samples while still requiring more
  samples (than plain dense subgraphs) for a high accuracy score. For all motifs,
  we set our minimal amount of training samples to 20,000 for positive and
  for negative ones.

\section*{Appendix D: Analysis of Different Variants of Motif Prediction in SEAM}

Here, we analyze the effects and contributions from different variants of SEAM.
First, we investigate the accuracy improvements due to our
proposed labeling scheme in Section~\ref{sec:labeling}. Then, we
empirically justify our approach to only sample the $h$-hop enclosing subgraph
for small $h$ (1--2). Finally, we evaluate the performance of every prediction
  method if there are no motif edges already present.

\vspaceSQ{-0.5em}
\subsection*{Labeling Scheme vs.~Accuracy}

Figure~\ref{fig:labels-usair} shows that our proposed labeling scheme 
generally has a positive impact on the accuracy of SEAM. The exception is
the {$k$-star} motif. For $k=3$, the labeling scheme significantly improves
the accuracy. On the other hand, using $k>3$ reduces the accuracy 
while simultaneously increasing the variance of test results. This
effect can be explained with the implementation details of our labeling scheme.
We remove every edges between all the motif vertices to calculate our
{$k$-dimensional} distance labels. This procedure seems to misrepresent
the structure of {$k$-stars} for $k>3$. There are possible improvements to
be gained in future work by further optimizing our labeling
scheme.

\iftr
\begin{figure*}[t]
\else
\begin{figure}[h]
\fi
\vspaceSQ{-1em}
\iftr
\centerline{\includegraphics[width=1.5\columnwidth]{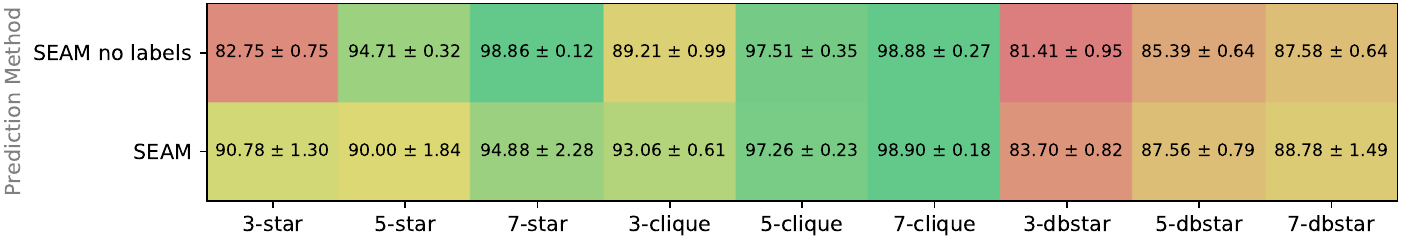}}
\fi
\vspaceSQ{-1em}
\caption{\textmd{Effect of our proposed labeling scheme on \textbf{USAir} graph. $h$-hop = 1, learning rate = 0.002, number of epochs = 100, 
%
%
training dataset size = 100,000.}}
\label{fig:labels-usair}
\vspaceSQ{-1em}
\iftr
\end{figure*}
\else
\end{figure}
\fi

\vspaceSQ{-0.5em}
\subsection*{$h$-Hop Enclosing Subgraphs vs.~Accuracy}

Zhang et al.~\cite{zhang2018link} motivated the use of small $h$-hop
neighborhoods for SEAL with the $\gamma$-decaying heuristic.
We now provide additional data to backup this decision in SEAM.
Figures~\ref{fig:hop-power} and \ref{fig:hop-usair} show that in most cases
there is not much performance to be gained by sampling an $h$-hop enclosing
subgraph with $h>2$. This effect is especially striking for sparse graph
datasets like the Power shown in Figure~\ref{fig:hop-power}. The accuracy
starts to drop significantly for $h>2$. The only outlier in our little test was
the 5-star motif shown in Figure~\ref{fig:hop-usair}. This effect was most
likely caused by the specifics of this particular dataset and it does reflect a
trend for other graphs. An additional explanation could also be the
non-optimal labeling implementation for the 5-star motif. These special
cases do not justify to increase the neighborhood size of the motif in a
general case.

\iftr
\begin{figure*}[t]
\else
\begin{figure}[h]
\fi
\iftr
\centerline{\includegraphics[width=1.2\columnwidth]{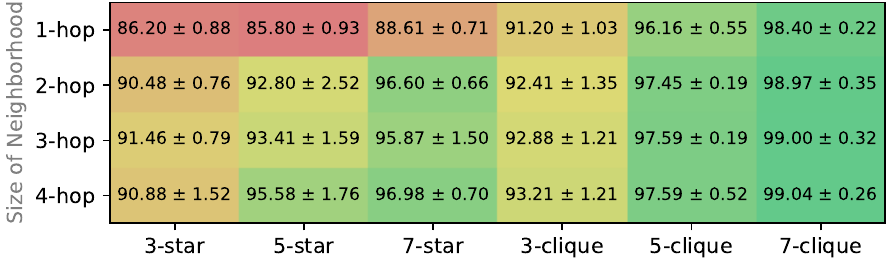}}
\fi
\vspaceSQ{-1em}
\caption{\textmd{Comparison of different $h$-hop enclosing subgraphs used in SEAM, for the \textbf{USAir} graph.
Learning rate = 0.002, number of epochs = 100.}}
%
%
\label{fig:hop-usair}
\vspaceSQ{-1em}
\iftr
\end{figure*}
\else
\end{figure}
\fi

\iftr
\begin{figure*}[t]
\else
\begin{figure}[h]
\fi
\iftr
\centerline{\includegraphics[width=1.2\columnwidth]{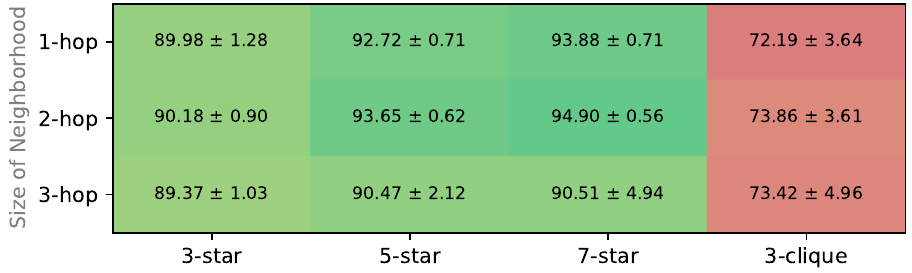}}
\fi
\vspaceSQ{-1em}
\caption{\textmd{Comparison of different $h$-hop enclosing subgraphs used in SEAM, for the \textbf{Power} graph.
Learning rate = 0.002, number of epochs = 100, training dataset size = 100,000.
%
%
The graph does not contain enough {5-cliques} and {7-cliques} due to its sparsity.}}
\label{fig:hop-power}
\vspaceSQ{-1em}
\iftr
\end{figure*}
\else
\end{figure}
\fi

\subsection*{Presence of Motif Edges vs.~Accuracy}

\iftr
\begin{figure*}[t]
\else
\begin{figure}[h]
\fi
\iftr
\centerline{\includegraphics[width=1.2\columnwidth]{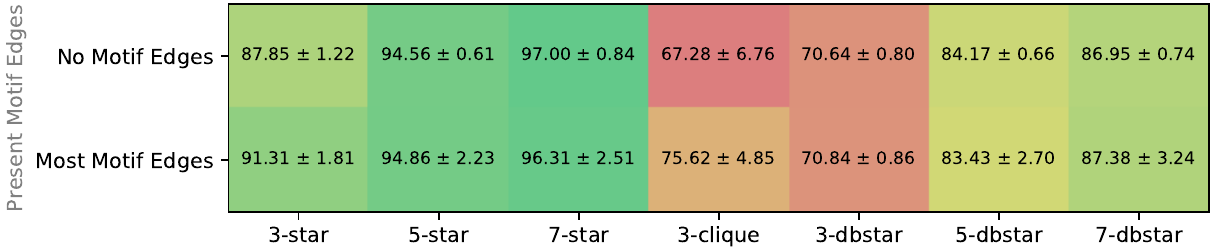}}
\fi
\vspaceSQ{-1em}
\caption{\textmd{Comparison of the prediction accuracy of SEAM for different already present motif edges for the \textbf{Power} graph.
$h$-hop = 1, learning rate = 0.002, number of epochs = 100, training dataset size = 100,000.
The graph does not contain enough {5-cliques} and {7-cliques} due to its sparsity.}}
%
%
\label{fig:no-motif-edges-power}
\vspaceSQ{-1em}
\iftr
\end{figure*}
\else
\end{figure}
\fi

\iftr
\begin{figure*}[t]
\else
\begin{figure}[h]
\fi
\vspaceSQ{-1em}
\iftr
\centerline{\includegraphics[width=1.5\columnwidth]{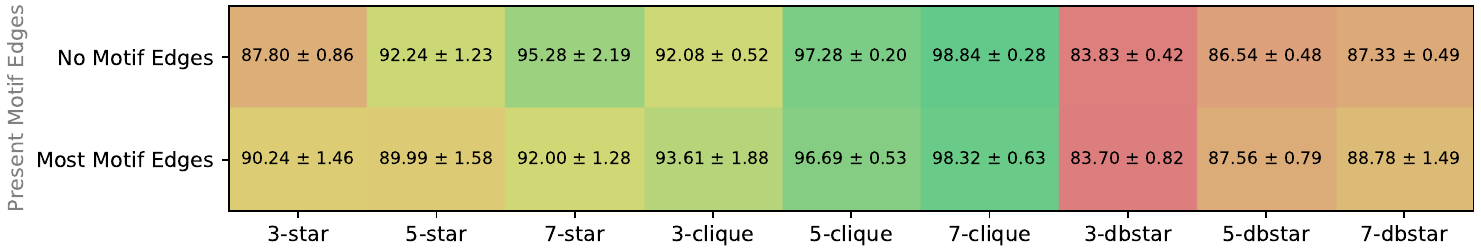}}
\fi
\vspaceSQ{-1em}
\caption{\textmd{Comparison of the prediction accuracy of SEAM for different already present motif edges for the \textbf{USAir} graph.
$h$-hop = 1, learning rate = 0.002, number of epochs = 100,  
%
%
training dataset size = 100,000.}}
\label{fig:no-motif-edges-usair}
\vspaceSQ{-1em}
\iftr
\end{figure*}
\else
\end{figure}
\fi

We now illustrate that SEAM also ensures high accuracy when \emph{no}
or \emph{very few} motif edges are already present, see Figures
\ref{fig:no-motif-edges-power} and \ref{fig:no-motif-edges-usair}. Thus, we can
conclude that SEAM's prediction strength relies mostly on the structure of the
neighborhood subgraph, embeddings, vertex attributes, and our proposed labeling
scheme, and not necessarily on whether a given motif is already partially
present. Outliers in this experiment are the {3--clique} in the Power graph,
the {$k$-star} motif with $k>3$ in the USAir graph, and the {3-star} motif in
general. Still, there is no general tendency indicating that SEAM would profit
greatly from the presence of most motif edges.

\section*{Appendix F: Details of Implementation \& Used Hardware}

Our implementation\footnote{\scriptsize Code will be available at
\url{http://spcl.inf.ethz.ch/Research/Parallel_Programming/motifs-GNNs/}} 
%
of SEAM and SEAL use the PyTorch Geometric Library~\cite{fey2019geometric}. We
employ Ray~\cite{moritz2017ray} for distributed sampling and preprocessing, and
RaySGD for distributed training and inference.

To run our experiments, we used the AULT cluster and the Piz Daint cluster at
CSCS~\cite{CSCS2021}. For smaller tasks, we used nodes from the AULT cluster
such as AULT9/10 (64 AMD EPYC 7501 @ 2GHz processors, 512 GB memory and 4
Nvidia V100 GPUs), AULT23/24 (32 Intel Xeon 6130 @ 2.10GHz processors, 1.5TB
memory and 4 Nvidia V100 GPUs), and AULT25 (128 AMD EPYC 7742 @ 2.25GHz
processors, 512 GB memory and 4 Nvidia A100 GPUs). For larger, tasks we used our
distributed implementation on the Piz Daint cluster (5704 compute nodes, each
with 12 Intel Xeon E5-2690 v3 @ 2.60GHz processors, 64 GB memory and a Nvidia
Tesla P100 GPU).

\end{document}